\documentclass[a4paper,11pt]{article}

\usepackage{amsmath,amssymb}           % amstex
\usepackage{amscd}                     % ams commutative diagrams
\usepackage{epsfig}                    % ps figures
\usepackage[matrix,arrow]{xy}          % commutative diagrams
\usepackage{xspace}                    % space in abbreviations
\usepackage{stmaryrd}                  % extra symbol font (dbl bracket)
\usepackage{slashed}
\usepackage{tabularx}
\usepackage{bbold}
\usepackage{jheppub}                   % jhep style
\makeatletter
\gdef\@fpheader{\ }                    % hack the jhep header
\makeatother

\newcommand{\half}{{{\textstyle\frac{1}{2}}}}

\newcommand{\beq}{\setlength\arraycolsep{2pt}\begin{equation}}
\newcommand{\eeq}{\end{equation} }
\newcommand{\bpm}{\begin{pmatrix}}
\newcommand{\epm}{\end{pmatrix}}
\newcommand{\mtr}[1]{\begin{pmatrix} #1 \end{pmatrix}} 

\newcommand{\comm}[2]{\left[#1,#2\right]}

\newcommand{\proj}[1]{\lbrack\!\lbrack #1 \rbrack\!\rbrack}
\newcommand{\order}[1]{\scriptscriptstyle (#1)}
\newcommand{\inv}[1]{#1^{\scriptscriptstyle -1}}

\newcommand\dd{{\rm d}}

\newcommand\cD{{\cal D}}
\newcommand\cE{{\cal E}}

\newcommand\cH{{\cal H}}

\newcommand\cJ{{\cal J}}

\newcommand\cL{{\cal L}}

\newcommand\cO{{\cal O}}
\newcommand\cP{{\cal P}}

\newcommand\cR{{\cal R}}
\newcommand\cS{{\cal S}}

\newcommand\hcL{{\hat{\cal L}}}

\newcommand\ha{\hat{a}}
\newcommand\hb{\hat{b}}
\newcommand\hc{\hat{c}}
\newcommand\hd{\hat{d}}
\newcommand\he{\hat{e}}
\newcommand\hf{\hat{f}}
\newcommand\hg{\hat{g}}

\newcommand\hI{\hat{I}}
\newcommand\hJ{\hat{J}}
\newcommand\hK{\hat{K}}
\newcommand\hL{\hat{L}}

\newcommand\hT{\hat{T}}

\newcommand\hX{\hat{X}}

\newcommand\hcH{\hat{\cH}}
\newcommand\hcJ{\hat{\cJ}}
\newcommand\hcO{\hat{\cO}}

\newcommand\hpartial{\hat{\partial}}

\def\tT{{\tilde{T}}}

\def\tf{\tilde{f}}

\def\tx{\tilde{x}}

\def\tz{\tilde{z}}

\def\tpartial{\tilde{\partial}}

\newcommand{\Odd}{\mathbf{O}(d,d)}

\def\tPi{\tilde{\Pi}}

\newcommand\hcHp{\hat{\cH}_{\scriptscriptstyle \Pi}{}}
\newcommand\cPo{\cP^{\scriptscriptstyle 0}_d}
\newcommand\sPi{{\scriptscriptstyle \Pi}}

\newcommand{\OddZ}{\mathbf{O}(d,d;\mathbb{Z})}

\begin{document}
\begin{titlepage}
\begin{flushright}
Q15005
\end{flushright}

\vfill

\begin{center}
   \baselineskip=16pt
   {\LARGE \bf Towards Weakly Constrained Double Field Theory 
    \\~\\~}
\\~\\
 \bf Kanghoon Lee \footnote{\tt kanghoon@kias.re.kr}
       \vskip .6cm
             \begin{small}
             	\vspace{2mm}
		{\it Korea Institute for Advanced Study, Seoul 130-722, Korea. }\\
\vspace{2mm}
\end{small}
\end{center}

\vfill 
\begin{center} 
\textbf{Abstract}
\end{center} 
\begin{quote}
We show that it is possible to construct a well-defined effective field theory incorporating string winding modes without using strong constraint in double field theory. We show that X-ray (Radon) transform on a torus is well-suited for describing weakly constrained double fields, and any weakly constrained fields are represented as a sum of strongly constrained fields. Using inverse X-ray transform we define a novel binary operation which is compatible with the level matching constraint. Based on this formalism, we construct a consistent gauge transform and gauge invariant action without using strong constraint. We then discuss the relation of our result to the closed string field theory. Our construction suggests that there exists an effective field theory description for massless sector of closed string field theory on a torus in an associative truncation.

\end{quote} 
\vfill
\setcounter{footnote}{0}
\end{titlepage}
\newpage
\tableofcontents 
%%%%%%%%%%%%%%%%%%%%%%%%%%%%%%%%%%%%%%%%%%%%%%%%%%%%%%%%%%%%%%%%%%%%%%%%%%%%%%%%%%%%%%%%%%%
\section{Introduction}

Recent progress in understanding T-duality through double field theory (DFT) \cite{Siegel:1993th,Siegel:1993xq,Hull:2009mi,Hull:2009zb,Hohm:2010jy,Hohm:2010pp} has lead to interest in constructing an effective theory describing string winding states. When a string is moving on a torus bundle where the radius of torus is near self-dual radius $R \simeq \sqrt{\alpha'}$, the momentum and winding states are treated symmetrically due to the T-duality. If we focus on the torus fibre, string states are specify by momentum $p$ and winding number $w$, hence target spacetime fields are also depend on both $p$ and $w$, or $x$ and $\tx$ which are the periodic coordinates for torus and its dual torus respectively. Such fields are called \emph{double fields} and defined on doubled tori. 

However, double fields are not arbitrary functions with respect to the $x$ and $\tx$, but they are constrained by level matching constraint
\begin{equation}
  \big(L_0 - \bar{L}_0\big) \Phi(x,\tx) = 0\,.
\label{}\end{equation}
For simplicity, if we consider only massless subsector $N=\bar{N}=1$, then the level matching constraint reduces to  
\begin{equation}
  \partial_I\partial^I \Phi(x,\tx) = 0\,.
\end{equation}
anthe massless fields are defined on a $(2d -1)$-dimensional cone in a doubled momentum space. The effective field theory for massless double fields is called double field theory. It is expected that DFT provides an effective field theory for closed strings on a torus background beyond the conventional supergravities. 

However, the full DFT has not been constructed yet. The main obstruction in constructing DFT is that the ordinary product of double fields $f(x,\tx)$ and $g(x,\tx)$ does not satisfy level matching constraint again
\begin{equation}
  \partial_I \partial^I (f \cdot g) \neq 0\,.
\label{violationLMC}\end{equation}
In order to make it possible to satisfy level matching constraint, we require so-called strong constraint that all the fields and the gauge parameters as well as all of their products should be annihilated
\begin{equation}
  \partial_I f \cdot \partial^I g = 0\,,
\end{equation}
and the strongly constrained fields are defined on a maximal null plane which is specified by section condition. Any field satisfying the strong constraint is called a strongly constrained field. By imposing the strong constraint, a consistent gauge transform and gauge invariant action has been constructed in a $\Odd$ covariant form \cite{Hohm:2010pp}. We will refer to the DFT with fields obeying the strong constraint as strongly constrained DFT to distinguish it from DFT with weakly constrained fields which we will call simply DFT if the strong constraint is not imposed. 

Another important issue for weakly constrained DFT is that string massive states cannot be decoupled near self-dual radius. An effective field theory includes the appropriate degrees of freedom to describe physical phenomena occurring at a given energy scale, while ignoring degrees of freedom at shorter distances. However, we cannot keep only massless states in closed string field theory near self-dual radius. In order to get a theory for massless degrees of freedom, we should integrate out all the massive fields by hand. Obviously, DFT is not a usual low energy effective field theory. Such computation is not practically possible, and it is not clear whether effective field theory description is valid. 

 Recent works have addressed the relaxation of the strong constraint in generalized Scherk-Schwarz reduction \cite{GSS1,GSS2,GSS3,GSS4,GSS5,GSS6}. It turns out that the generalize twist matrix or generalized frame fields are not necessary to satisfy level matching constraint. However, it has been shown that if we relax the strong constraint, then the weak constraint is also violated \cite{GSS2}. It is not clear whether such backgrounds are well-defined as string backgrounds. It is known for example that such backgrounds violate modular invariance \cite{Lee:2015xga}.

In the present paper, we show that a full relaxation of strong constraint is possible, and we explicitly construct a well-defined gauge transform and associated gauge invariant action without using the strong constraint. The main ingredient for this construction is the X-ray (or Radon) transform on a torus \cite{RadonTorus1,RadonTorus2,RadonTorus3}. Usually it is applied to X-ray images in tomography. In the context of DFT, the X-ray transform is used to represent a weakly constrained field on a doubled torus in terms of strongly constrained fields which are defined on all possible null-planes. In fact, this idea is closely related to the Penrose transform in twistor the formalism \cite{Penrose:1969ae,Penrose:1977in}. There is a remarkable correspondence between DFT and the Penrose transform. In the Penrose transform, massless fields are represented by a sum of fields defined on all possible light-cones. If we take our metric signature as $\mathbf{O}(2,2)$, the wave equation for massless fields can be identified to level matching constraint. The light-cone corresponds to the maximal null subspace, and it defines the so called {\it{section condition}}. Since the X-ray transform is a real version of Penrose transform, the X-ray transform is well-suited to describe weakly constrained fields.
\begin{table}
\begin{center}
\begin{tabular}{c|c}
  DFT & Penrose transform \\
  \hline
  weakly constrained fields & massless fields\\
 level matching constraint & wave equation  \\
 strong constraint & light cone
\end{tabular}
\caption{Comparison between DFT and Penrose transform}
\end{center}
\end{table}

We give a prescription to resolve the issue discussed in (\ref{violationLMC}) and define a novel binary operation, $\circ$, which is compatible with the level matching constraint, through X-ray transform. For weakly constrained fields $f$ and $g$, the $f\circ g$ satisfies level matching constraint
\begin{equation}
  \partial_I \partial^I \big(f \circ g\big) = 0\,,
\end{equation}
and a strong constraint like identity
\begin{equation}
  \partial_I f \circ \partial^I g = 0\,.
\end{equation}
In addition, the $\circ$-product satisfies the commutative, associative and other useful algebraic properties. This framework is both practical and conceptual. We also discuss the relation between $\circ$-product and Hull and Zwiebach's projector \cite{Hull:2009mi} which originated from the string product in closed string field theory. The HZ projector is also compatible with the level matching constraint and is commutative, but it is not associative. We show that the difference between $\circ$-product and HZ projector arises from zero-modes in Fourier expansion of weakly constrained fields. Hence, if there exists an appropriate physical truncation which eliminates all just the zero-modes, then the HZ projector reduces to $\circ$-product and defines an associative truncation. However, it is not clear what kind of physical truncation eliminates the Fourier zero-modes only. 

The physical degrees of freedom are represented by a weakly constrained generalized metric and a dilaton which consist of a set of strongly constrained generalized metrics and dilatons respectively. We introduce T-duality transform for weakly constrained fields incorporating the $\circ$-product, denoted by $\OddZ_\circ$, and show that it forms a well-defined group. We show that the weakly constrained generalized metric and dilaton are \(\OddZ_\circ\) tensor and scalar respectively. Using the fact that the weakly constrained fields are represented by a sum of strongly constrained fields, we define a gauge transform for weakly constrained DFT as a sum of generalized Lie derivatives, which is the gauge transform of strongly constrained DFT. Also, we show that the gauge symmetry has a closed gauge algebra without strong constraint. We then construct an action for weakly constrained DFT which is invariant under the gauge transform. By a similar argument as the gauge transform, this action is represented by a sum of all possible strongly constrained DFT actions. 

Although we have constructed a consistent gauge transform and gauge invariant action without using the strong constraint, the relation to closed string field theory remains an open question. We show that up to cubic order of fluctuations our action and closed string field theory action around the same constant background are not equivalent. This disagreement arises from the difference between $\circ$-product and HZ projector. As already mentioned, weakly constrained DFT is assumed to be obtained by integrating out all the massive string states on a near self-dual torus even though we cannot decouple the string massive states. In general such a calculation is practically impossible, and it is not clear what is the consistent fluctuations for closed string field theory in this situation. Since $\circ$-product provides a certain sort of associative truncation of the HZ projector, under this truncation the disagreement between our result and closed string field theory disappears. Hence, at least our construction may provide some sort an associative trauncation for the massless subsector of closed string field theory on a torus.

The organization of the present paper is as follows. In section 2, we review the X-ray transform on a doubled torus and inverse X-ray transform.  We show that weakly constrained fields are reconstructed by a sum of strongly constrained fields. Using the inverse X-ray transform we define a binary operation, $\circ$, which is compatible with the level matching constraint. We also compare the $\circ$-product with Hull and Zwiebach's projector. In section 3, we define $\OddZ$ T-duality transform equipped with the $\circ$-product for weakly constrained fields. The weakly constrained generalized metric and dilaton are introduced and identified as the physical degrees of freedom.  We end in section 4 by constructing a gauge transform and gauge invariant action for weakly constrained DFT. We also discuss the relation with the closed string field theory result.

%%%%%%%%%%%%%%%%%%%%%%%%%%%%%%%%%%%%%%%%%%%%%%%%%%%%%%%%%%%%%%%%%%%%%%%%%%%%%%%%%%%%%%%%%%%
\section{X-ray transform}
%%%%%%%%%%%%%%%%%%%%%%%%%%%%%%%%%%%%%%%%%%%
\subsection{Closed $d$-planes}\label{Sec:d-plane}
Before considering the X-ray transform, we will review some basic facts about $d$-dimensional closed plane on a doubled torus $T^{2d}$ \cite{RadonTorus1,RadonTorus2,RadonTorus3}. We introduce $2d$-periodic coordinates for the $T^{2d}$, $X^I = (x^i,\tx_{i})$, which are identified according to
\begin{equation}
  x^i \sim x^i + 1  \,, \qquad \tx_i \sim \tx_i + 1 \,.
\end{equation}
Here $I,J, \cdots$ are $\Odd$ vector indices, and they are raised and lowered with the $\Odd$ metric $\cJ$
\begin{equation}
  \cJ_{IJ} = \bpm 0 & \delta^{i}{}_{j} \\ \delta_{i}{}^{j} & 0\epm\,.
\end{equation}

 A closed $d$-dimensional plane $\cD(X^I,\Pi)$ on a $T^{2d}$ passing through a point $X^I\in T^{2d}$ is parametrized as
\begin{equation}
  \cD(X^I,\Pi) = \{X^I+t_i \Pi^{iI}|  0\leq t_i < 1~ \text{and} ~ \Pi \in \cP_d\}\,.
\label{dplane} \end{equation}
For the periodicity of the coordinates $X^I$, the range of the parameters $t_i$ are given by $0$ to $1$. The $\cP_d$ is a set of $d\times 2d$ integer matrices of rank $d$, whose Smith normal form is given by 
\begin{equation}
   \Pi = L D_0 V\,,
\end{equation}
where $L \in PSL(d,\mathbb{Z})$, $V\in PSL(2d,\mathbb{Z})$ and $D_0 = (\mathbf{1}_{d} ~0_{d})$. This can be understood as the diagonalization of non-square matrices with unit eigenvalues. Then, the row vectors for $\Pi \in \cP_d$ are linearly independent and all the components of any row vectors are coprime \cite{},
\begin{equation}
  \gcd(\Pi^i) = 1\,, \qquad 1 \leq i \leq d\,,
\label{}\end{equation}
where the $\gcd(\Pi^i)$ is the greatest common divisor of $\Pi^i = (\Pi^i{}_1,\Pi^i{}_2, \cdots , \Pi^i{}_{2d} ) $.
Since the closed $d$-dimensional plane is represented as a section or cutting plane of $T^{2d}$, and the $\Pi$ determines how to slice, we will call the $\Pi$ as a \emph{slicing matrix}. The row vectors $\Pi^i$ are tangent vectors of the $\cD(X,\Pi)$.

For later use we also present a parametrization of a $d$-dimensional closed \emph{null} plane:
\begin{equation}
  \cD^{\scriptscriptstyle 0}(X^I,\Pi) = \{X^I+t_i \Pi^{iI}|  0\leq t_i < 1~ \text{and} ~ \Pi \in \cPo\}\,,
\label{nulldplane} \end{equation} 
where the $\cPo$ is a subset of the $\cP_d$ whose row vectors of $\Pi \in \cPo$ are null and mutually orthogonal
\begin{equation}
  \Pi^i{}_I \cJ^{IJ} (\Pi^t)_J{}^j = 0\,.
\label{nullPi}\end{equation}
Then the Smith normal form of the $\Pi\in \cP^0_d$ is given by
\begin{equation}
  \Pi = L D_0 V
\end{equation}
where $L \in PSL(d,\mathbb{Z})$ and $V\in \OddZ$. It is straightforward to show that any $\Pi\in \cPo$ satisfies the null condition (\ref{nullPi})
\begin{equation}
  \Pi^i{}_I \cJ^{IJ} (\Pi^t)_{J}{}^j = L D_0 V \cJ V^t D^t_0 L^t = L D_0 \cJ D_0 L^t = 0\,,
\end{equation}
as we expected.

Note that the parametrization (\ref{dplane}) and (\ref{nulldplane}) are not unique, but there is a $PSL(d,\mathbb{Z})$ equivalence class 
\begin{displaymath}
  \Pi^i{}_I  \sim a^i{}_j \Pi^j{}_{I}\,, \qquad a^{i}{}_{j} \in PSL(d,\mathbb{Z})\,.\end{displaymath}
In other words, if two slicing matrices $\Pi'$ and $\Pi$ are related by $PSL(d,\mathbb{Z})$ rotation, then they parametrize the same $d$-plane because the $a\in PSL(d,\mathbb{Z})$ can be absorbed into the parameter $t^i$ by redefining $t'_i = t_j a^j{}_i$.

%%%%%%%%%%%%%%%%%%%%%%%%%%%%%%%%%%%%%%%%%%%%%%%%%%%
\subsection{X-ray transform of weakly constrained fields}
Now we consider $d$-dimensional X-ray (or Radon) transform. The X-ray transform is an integral transform mapping a continuous function $f$ on a torus $T^{2d}$ to the integrals of this function over the $d$-dimensional closed planes $\cD(X^I, \Pi)$\footnote{X-ray (Radon) transform can be extedned to any $n$-dimensional closed planes, where $1\leq n\leq 2d-1$, however, we will focus only on the $n=d$ case. }
\begin{equation}
  \cR f({X}^I;\Pi) = \int_{0}^{1}\cdots\int_{0}^{1} \dd t_1 \cdots \dd t_d f\big(X^I + t_i \Pi^i{}^{I}\big)
\label{Xray}
\end{equation}
where $ X^I$ is a point on the $T^{2d}$ and $\Pi^{iI} \in \cP_d$. This means averaging the $f$ over the given $d$-dimensional plane $\cD(X^I,\Pi)$. Thus $\cR f(X^I;\Pi)$ has a translational invariance along the tangential direction of the plane. One of the remarkable aspects of the X-ray transform is that it is an injective mapping \cite{RadonTorus1,RadonTorus2,RadonTorus3}. Thus it is possible to define the inverse transform for a given plane. We will discuss the inverse X-ray transform in the next subsection. In general X-ray transform can be applied for any continuous functions on $T^{2d}$. However, from now on we will focus on weakly constrained fields, which satisfy
\begin{equation}
  \partial_I \partial^I f = 0\,. 
\end{equation}
 
To make the X-ray transform tangible, let us consider a null plain wave $e_K = e^{2\pi iK_IX^I}$ on a $T^{2d}$ with a $2d$-dimensional momenta $K_I\in \mathbb{Z}^{2d}$ satisfying
\begin{equation}
  K_I K^I = 0\,,
\label{nullK}\end{equation}
so that the $e_K$ satisfy level matching constraitnt.
Then the $t$ integrals in X-ray transform (\ref{Xray}) for the $e_K$ can be done trivially
\begin{equation}
\begin{aligned}
  \cR \,e_K(X^I;\Pi) & = \int \dd^d t \,e^{2\pi i K_I (X^I + t_i \Pi^{iI})} =  e^{2\pi i K_I X^I} \int \dd^d t\, e^{2\pi i K_I  t_i \Pi^{iI}}
  \\
  & = e_K\, \delta_{\Pi^{iI}K_I, 0} 
\end{aligned}
\label{XrayFourier}\end{equation}
Then, for a given slicing matrix $\Pi$, the $K_I$ is constrained by two conditions:
\begin{equation}
\begin{aligned}
  \Pi^{iI} K_I = 0\,, \qquad & (1)
  \\
  K_I \cJ^{IJ} K_J = 0\,. \qquad &(2)	
\end{aligned}\label{conditionsK}
\end{equation}

Now let us interpret this result. For every momentum $K_I$, the $d$-equations from the first condition in (\ref{conditionsK}),
\begin{equation}
  \Pi^{iI} K_{I} = 0\,, \qquad i = 1,2, \cdots,d
\label{momentaConst1}
\end{equation}
eliminate the $d$-degrees of freedom among the $2d$ components of $K^I$, and there are $d$ remaining independent components. Let us denote the $d$ independent component as $\ell_i$. Then the $K_I$ is given by as a linear combination of the $\ell_i$ 
\begin{equation}
  K_I = \ell_i \Psi^i{}_I\,,
\label{expansionK}\end{equation}
where the $\Psi$ is a $d\times2d$ integer valued matrix of rank $d$. From the second condition in (\ref{conditionsK}), the $\Psi^i$ are mutually null and orthogonal vectors, 
\begin{equation}
  \Psi^i{}_{I} \cJ^{IJ} \Psi^{i}{}_{J} = 0\,,
\end{equation}
then the row vectors $\Psi^i$ become a basis of a maximal null subspace $N$. Also, the $\Pi^i$ and $\Psi^i$ are orthogonal to each other by the constraint (\ref{momentaConst1})
\begin{equation}
  \Pi^i{}_{I} \cJ^{IJ} \Psi^{j}{}_{J} =0\,.
\end{equation}
Recall that the orthogonal complement of a maximal null subspace $N$ is identical with itself, $N=N_{\perp}$. Since the $\Pi^i$ generates the $N_{\perp}$, we can identify $\Pi$ and $\Psi$ without loss of generality. After the identification  (\ref{expansionK}) is replaced as
\begin{equation}
  K_I = \ell_i \Pi^i{}_I\,,
\label{def_l}\end{equation} 
and one can show that the $\Pi^i$ are null vectors from the second condition in (\ref{conditionsK}) 
\begin{equation}
  \Pi^i{}_{I} \cJ^{IJ} \Pi^{i}{}_{J} = 0\,.
\label{orthogPi}
\end{equation}

As we have defined in (\ref{nulldplane}), the slicing matrix $\Pi$ defines a null $d$-dimensional plane $\cD^0(X^I,\Pi\in\cPo)$ because the tangent vectors of the $D(X^I, \Pi)$ are orthogonal
\begin{equation}
  \frac{\partial}{\partial X^I} \cJ^{IJ} \frac{\partial}{\partial X^J} = \Pi^i{}_{I} \Pi^{jI} \frac{\partial}{\partial z^i} \frac{\partial}{\partial z^j} = 0\,.
\end{equation}
This shows that the X-ray transform of a null plane wave makes sense only on null-planes. The X-ray transform of the $e_K$ (\ref{XrayFourier}) can be rewritten in terms of the $d$-dimensional momenta $\ell_i$ 
\begin{equation}
  \cR e_K(X^I;\Pi^i) = e^{2\pi i \ell_{i}\Pi^i{}_I X^I} = e^{2\pi i \ell_{i}z^i}\,,
\label{XrayPlaneWave}
\end{equation}
where $z^i = \Pi^{i}{}_{I}X^{I}$. The $z^i$ are coordinates for a $d$-dimensional null-plane. Then X-ray transform of a Fourier basis $e_K$ on $T^{2d}$ reduces to a $d$-dimensional Fourier basis on $d$-dimensional null plane defined by $\Pi^i{}_{I}$. 

In order to extend the X-ray transform to an arbitrary weakly constrained field, we carry out Fourier expansion with respect to the $X^I$ and use the result of X-ray transform of the null plane wave (\ref{XrayPlaneWave})
\begin{equation}
\begin{aligned}
  \cR f(X^I;\Pi^i) = & 
	\sum_{K \in \mathbb{Z}^{2d}} \tf_{K} e^{2\pi i K_I X^I}	 \delta_{\Pi^{iI} K_I, 0}
\\
=& \sum_{l_i} \tf_{\ell_i \Pi^i}\, e^{2\pi i l_i z^i}\,.
\end{aligned}\label{FourierSlice}
\end{equation}
This is the usual Fourier expansion on a $d$-dimensional null plane. It is known as \emph{Fourier slice theorem}. Since strongly constrained fields are defined on null-planes which are defined by section condition, therefore, 
\begin{quotation}
\emph{The X-ray transform maps a $2d$-dimensional weakly constrained field to a $d$-dimensional strongly constrained field on a $d$-dimensional null plane.}
\end{quotation}

%%%%%%%%%%%%%%%%%%%%%%%%
\subsection{Inverse X-ray transform for weakly constrained fields}\label{sec:InverseXray}
In the previous subsection, we have shown that X-ray transform maps a $2d$-dimensional weakly constrained field $f(X^I)$ to a strongly constrained $d$-dimensional function $\cR f(X^I;\Pi)$ on a null $d$-plane $\cD^0(X;\Pi)$. As we discussed before, since the X-ray transform is an injective mapping, one can define an inverse X-ray transform. For a given null-plane specified by $\Pi$, inverse X-ray transform reconstructs the original weakly constrained field $f$ in terms of its X-ray images $\cR f(X^I;\Pi)$. 

As the X-ray transform can be applied to any continuous functions on a $T^{2d}$ regardless of the level matching constraint, the inverse formula is also defined for an arbitrary continuous function $f$ \cite{RadonTorus1,RadonTorus2,RadonTorus3}
\begin{equation}
\begin{aligned}
  f(X^I) & = \sum_{K_I\in\mathbb{Z}^{2d}} \tfrac{1}{\psi(K)} \int_{T^{2d}} \dd^{2d}\, Y  e^{2\pi i K_I(X^I-Y^I)} \sum_{\Pi\in\cPo} \varphi(\Pi) \cR f (\Pi^i{}_I Y^I) 
\end{aligned}\label{inverseXray}
\end{equation}
where $\varphi(\Pi^i)$ is a weight factor which ensures the convergence of the series
\begin{equation}
  \varphi(\Pi^i) = \exp(-\| \Pi^i{}_I\|^2 ) = \exp(- \sum_{i,I} (\Pi^i{}_{I})^2)\,,
\end{equation}
and the $\psi(K)$ is a normalization factor associated with $\varphi(\Pi)$
\begin{equation}
  \psi(K;\Pi^i) = \sum_{\Pi^i\in\cP_d \atop \Pi^i{}_{I} K^I = 0} \varphi(\Pi^i)\,.
\label{psi}\end{equation}

In order to rewrite the inverse formula in the form where the meaning of the inverse transform is manifest, we introduce a $d$-dimensional field $\hat{f}_{\sPi}(z^i)$
\begin{equation}
  \hf_{\sPi}(z^i) = \int_{T^{2d}} \dd^{2d}\, Y  \sum_{K} \tfrac{1}{\psi(K)} \cR f(Y^I;\Pi) e^{2\pi i K_I(X^I-Y^I)}\,.
\label{hatOp}\end{equation}
This allows us to express the inverse formula in more intuitive way
\begin{equation}
  f(X^I) = \sum_{\Pi\in \cP_d} \varphi(\Pi) \hf_{\sPi}(z^i)\,.
\label{inverseXray}\end{equation}

Now let us assume that the $f$ is a weakly constrained field. After carrying out a Fourier expansion for the $\cR f(Y^I;\Pi)$ (\ref{FourierSlice}), one can show that the X-ray image fields $\hf_{\sPi}(z^i)$ reduce to
\begin{equation}
  \hf_{\sPi}(z^i) = \tfrac{1}{\psi(0)} \cR f(X^I;\Pi)\,.
\label{Xrayimage}\end{equation}
$\cR f(X^I;\Pi)$ are X-ray images of $f$, we will call the hatted fields $\hf_{\sPi}$  \emph{X-ray image fields}. Moreover, each $\hf_{\sPi}(z^i)$ are on a null plane $\cD^0(X^I,\Pi)$, thus they satisfy strong constraint due to the null-property of $\Pi^i$
\begin{equation}
  \partial_I \hf_{\sPi} \cdot \partial^I \hg_{\sPi} = 0\,.
\end{equation}
In the following the hatted quantities mean strongly constrained fields. Therefore, the expression (\ref{inverseXray}) yields that
\begin{quotation}
  \emph{Weakly constrained fields are represented as a sum of strongly constrained fields through inverse X-ray transform.}
\end{quotation}

As a simple check of the inversion formula (\ref{inverseXray}), let us consider the null plane wave $e_K = e^{2\pi i K_I {X}^I}$ again. Using the X-ray transform of $e_K$ (\ref{XrayFourier}), $\he_{K,\Pi}(z^i)$ is given by 
\begin{align}
  	\he_{K,\Pi}(z^i) &= \sum_{K'} \tfrac{1}{\psi(K')} e^{2\pi i K'_{I} {X}^I} \int_{T^{2d}} \dd^{2d}\, Y ~ e^{-2\pi i (K'_I - K_I) {Y}^I} \delta_{\Pi^i{}_I K^I,0} \nonumber
\\
	&= \sum_{K'} \tfrac{1}{\psi(K')} e^{2\pi i K'_I {X}^I} \delta_{K_I,K'_I} \delta_{\Pi^i{}_I K^I,0}
\\	
	&= \tfrac{1}{\psi(K)} e^{2\pi i K_I {X}^I} \delta_{\Pi^i{}_I K^I,0} \nonumber
\end{align}
If we substitute $\he_{K,\sPi}$ into the inverse X-ray transform (\ref{inverseXray}), then  we have
\begin{align}
  e_K = \sum_{\Pi\in\cPo} \varphi(\Pi) \tfrac{1}{\psi(K)} e^{2\pi i K_I {X}^I} \delta_{\Pi^i{}_I K^I,0}
\end{align}
Since $K = \ell_i \Pi^i$ is orthogonal to $\Pi^i$, then $\psi(K;\Pi^i)$ reduces
\begin{equation}
  \psi(K;\Pi^i) = \sum_{\Pi\in\cPo} \varphi(\Pi^i) \delta_{\Pi^i K,0} = \psi(0)\,.
\end{equation}
Thus, the inverse X-ray transform formula is verified for the plane wave $e_K = e^{2\pi i K_I X^I}$. This result can be extended to any weakly constrained fields through Fourier expansion. 

Another simple example is a constant $c$. The X-ray tranform of $c$ is independent of the slicing matrix $\Pi$,
\begin{equation}
 \cR c(X^I;\Pi) = c \,,
\label{Xrayconstant}\end{equation}
and the X-ray image of $c$ is simply $\hc_{\sPi} = \frac{1}{\psi(0)} c$. The inverse X-ray transform is trivially satisfied
\begin{equation}
  c = \sum_{\Pi\in\cPo} \varphi(\Pi) \tfrac{1}{\psi(0)} c\,.
\end{equation}
It is important to note that each X-ray image for a constant field is identical and independent to the $\Pi$. Thus inverse X-ray transform of the product with a constant field $c$ is
\begin{equation}
  c\cdot f(X^I) = \sum_{\Pi\in \cP^0_d} \varphi(\Pi) c \hf_{\sPi}(z^i)
\end{equation}
and the X-ray image is simply divided as
\begin{equation}
  \widehat{c\cdot f}_{\sPi} = c \cdot \hf_{\sPi}
\end{equation}
Note that the weight factor $\varphi(\Pi^i)$ is not unique at all, but it is introduced by hand for the convergence of the series. However, the final result is independ on the choice of $\varphi(\Pi^i)$ due to the normalization factor $\frac{1}{\psi(0)}$ in the X-ray image (\ref{Xrayimage}).

%%%%%%%%%%%%%%%%%%%%%%%%%
\subsection{Relation to Penrose transform}
Now we will discuss the relation between inverse X-ray transform and Penrose transform \cite{Penrose:1969ae,Penrose:1977in}. Let us consider flat $\mathbb{R}^{2d}$ with $\mathbf{O}(d,d)$ metric signature instead of torus background. In this case, the summation in the inverse X-ray transform (\ref{inverseXray}) is replaced by integration over all possible null $d$-dimensional planes. Using the $PSL(d;\mathbb{R})$ equivalence relation, we can remove the redundancy of the parametrization by fixing the slicing matrix $\Pi$ as \footnote{Since we are considering a non-compact space, the $PSL(d;\mathbb{Z})$ is replaced to $PSL(d;\mathbb{R})$.} \footnote{For a torus case, such a choice of slicing matrix is not allowed because $u^{ij}$ are not integer valued matrices in general.}
\begin{equation}
  \Pi = \mtr{\mathbf{1}_{d} & ~u^{ij}}
\end{equation}
where $u^{ij}$ is an antisymmetric $d\times d$ matrix due to the null property of the $\Pi$ (\ref{orthogPi}). Then the inverse X-ray transform is rewritten as
\begin{equation}
  f(X^I) = \int \dd u\, \hf_{u} (z^i)
\label{ContInvXray}\end{equation}
where the coordinate on a null $d$-plane $z^i$ is defined as
\begin{equation}
  z^i = x^i + u^{ij} \tx_j
\end{equation}
For $O(2,2)$ case, the $u^{ij}$ is $u^{ij} = \epsilon^{ij} u$.

Now let us consider the Penrose transform in the case of $O(2,2)$. In this metric signature, the twistor become a real two-component spinor. As we discussed in the introduction, a Penrose transform represents massless fields, which satisfy the Klein-Godon equation $\partial_I \partial^I \Phi=0$, in terms of fields on light cone. For a spin zero field the Penrose transform is given by as follows:
\begin{equation}
  f(X) = \oint \lambda_a \dd \lambda^a  F(\lambda,w)|_{w= X\lambda}\,,
\label{Penrosetransf}\end{equation}
where the constraint $w_{\dot{a}}= X_{a\dot{a}}\lambda^a$ is known as the incidence relation.
Using $O(2,2)$ rotation we can always choose a specific form for the twistor:
\begin{equation}
  \lambda^a = \mtr{1\\u}
\end{equation}
If we substitue this choice into the Penrose transform (\ref{Penrosetransf}) and use an appropriate gamma matrix representation, then we have
\begin{equation}
  f(X) = \int \dd u F(x^i + u^{ij} \tx_j)\,,
\end{equation}
and this is exactly same as inverse X-ray transform (\ref{ContInvXray}). Analogously, we can do similar analysis for arbitrary $\Odd$ case by using pure spinor \cite{Berkovits:2004bw}. Therefore, the X-ray transform is the real version of Penrose transform.

%%%%%%%%%%%%%%%%%%%%%%%%%
\subsection{Binary operation for weakly constrained fields}\label{Sec:convolution}
Suppose that $K$ is the kernel of level matching operator, $\partial_I \partial^I$. Then any weakly constrained field is an elements of the kernel $K$. As discussed in the introduction, even if $f$ and $g$ are weakly constrained fields, their usual product is not a weakly constrained field again
\begin{equation}
  \partial_I \partial^I (f\cdot g) = \partial_I f \cdot \partial^I g \neq 0\,.
\end{equation}
This implies that the $K$ is not closed by ordinary product
\begin{equation}
  f \cdot g \notin K\,.
\label{notclosed}\end{equation}
This is the main difficulty in constructing the weakly constrained DFT. For instance, the gauge transform of a weakly constrained field $\cH(x,\tx)$ would be a product of the gauge parameter $X(x,\tx)$ and the $\cH(x,\tx)$. However, their ordinary product is not a weakly constrained field, thus we cannot define a gauge transform using ordinary product. 
In order to construct a theory of weakly constrained fields, it is necessary to define a binary operation which is compatible with the level matching constraint
\begin{equation}
  f \circ g \in K.
\end{equation}

Let us assume that $f$ and $g$ are weakly constrained fields. From the inverse X-ray transform (\ref{inverseXray}), $f$ and $g$ are expanded by their X-ray images
\begin{equation}
  f(X^I) = \sum_{\Pi \in \cP^0_d} \varphi(\Pi) \hat{f}_{\scriptscriptstyle\Pi}(z^i;\Pi)\,, \qquad 
g(X^I) = \sum_{\Pi^{\prime} \in \cP^0_d} \varphi(\Pi) \hat{g}_{\scriptscriptstyle\Pi'}(z^{\prime}{}^{i};\Pi^{\prime})\,,
\label{inversXrayfg}\end{equation}
where the $z^{i} = \Pi^{i}{}_{I} {X}^{I}$ and $z^{\prime}{}^{i} = \Pi^{\prime}{}^{i}{}_{I} {X}^{I}$. Using (\ref{inversXrayfg}) $f\cdot g$ can be rewritten as
\begin{equation}
  f \cdot g = \sum_{\Pi,\Pi'\in\cPo} \varphi(\Pi) \varphi(\Pi') \hf_{\scriptscriptstyle \Pi}(z^i)  \hg_{\scriptscriptstyle \Pi'}(z'^i) \,.
\end{equation}
Here the X-ray image fields $\hf_{\scriptscriptstyle \Pi}$ and $\hg_{\scriptscriptstyle \Pi'}$ are defined on the two independent null planes, $\cD(X^I,\Pi)$ and $\cD(X^I,\Pi')$. As each X-ray image for a weakly constrained field is defined on a single plane, $f\cdot g$ cannot be a weakly constrained field. 

To investigate the reason for the violation of level matching constraint for the $f\cdot g$, we act the level matching operator $\partial_{I} \partial^{I}$ on $f\cdot g$. Short calculation using the chain rule shows that
\begin{equation}
  \partial_{I} \partial^{I} \big(f\cdot g\big) = 2 \partial_{I} f \cdot\partial^{I} g 
= 2 \sum_{\Pi, \Pi'\in \cPo} \varphi(\Pi) \varphi(\Pi') \Pi^{i}{}_{I} \Pi'{}^{jI}\, \frac{\partial \hat{f}_{\scriptscriptstyle\Pi}}{\partial z^i}  \frac{\partial\hat{g}_{\scriptscriptstyle\Pi'}}{\partial z'{}^j} \,,
\label{OrdiProd}\end{equation}
and it does not vanish in general as we expected.

A simple and natural way that makes the right-hand side of (\ref{OrdiProd}) vanish is to impose an orthogonality condition between $\Pi^i$ and $\Pi'^i$ \footnote{This condition is not the most general solution for the vanishing condition. We will discuss this issue in section \ref{sec:HZproj}.}
\begin{equation}
  \Pi^i{}_I \cJ^{IJ} \Pi'^{j}{}_{J} = 0\,.
\label{orthog}\end{equation}
Since the row vectors for a $\Pi$ generates a maximal null subspace, their orthogonal complement is identical with the original maximal null subspace generated by $\Pi^i$. Thus the $\Pi'^i$ is represented by a linear combination of $\Pi^i$
\begin{equation}
  \Pi'^i{}_I = a^i{}_j \Pi^j{}_I\,,
\end{equation}
where the $a^i{}_j$ is an element of $PSL(d;\mathbb{Z})$. Due to the $PSL(d;\mathbb{Z})$ equivalence relation, we can identify the $d$-dimensional null planes associated with the slicing matrices $\Pi$ and $\Pi'=a \Pi$,
\begin{equation}
  \cD^0(X^I;\Pi) = \cD^0(X^I; a\Pi)\,.
\end{equation}
This implies that the X-ray image fields $\hf_{\sPi}$ and $\hg_{\sPi'}$ are defined on the same null-plane. Moreover, we can absorb the $PSL(d;\mathbb{Z})$ matrix $a^i{}_j$ into the momenta $\ell_i$, which is introduced in (\ref{def_l}), by redefining $\ell'_i$ as
\begin{equation}
  \ell''_{i} = \ell'_j a^j{}_i\,.
\end{equation}
Therefore, we can always identify $\Pi$ and $\Pi'$ without loss of generality.

In accord with the remarks above, we define a novel binary operation $\circ$ for weakly constrained fields 
\begin{equation}
  f(X^I) \circ g(X^I) = \sum_{\Pi\in \cP^0_d} \varphi(\Pi) \hf{}_{\scriptscriptstyle\Pi}(z^i) \cdot \hat{g}{}_{\scriptscriptstyle \Pi}(z^i)\,.
\end{equation}
It is straightforward to check that $\circ$-product is compatible with level matching constraint due to the null property of the $\Pi$
\begin{equation}
  \partial_{I} \partial^{I} \big(f\circ g\big) = 0\,,\qquad \partial_I f \circ \partial^I g =0\,.
\label{level_matching_star_product}
\end{equation}

We can show that the $\circ$-product satisfies the following algebraic properties:
\begin{itemize}
	\item Commutativity
\begin{equation}
  f\circ g = g \circ f
\end{equation}
	\item Associativity
\begin{equation}
  f\circ(g\circ h) = (f\circ g)\circ h
\end{equation}
	\item Distributivity
\begin{equation}
  f\circ (g+h) = f\circ g + f \circ h  
\label{algebraicProperties}\end{equation}
\end{itemize}
Thus the kernel $K$ with $\circ$-product defines a commutative ring.

It is straight forward to define an identity $I$ satisfying $I\circ f = f \circ I = f$
\begin{equation}
  I = \sum_{\Pi\in \cPo} \varphi(\Pi) \cdot 1 = \psi(0)\,,
\end{equation}
where $\psi(0)$ is a constant defined in (\ref{psi}). It implies that for any two constants $a$ and $b$, their $\circ$-product is given by
\begin{equation}
  a\circ b= \sum_{\Pi\in\cPo}\varphi(\Pi) \ha_{\sPi} \cdot\hb_{\sPi}
\end{equation}
where $\ha_{\sPi}$ and $\hb_{\sPi}$ are given by
\begin{equation}
  \ha_{\sPi} = \tfrac{1}{\psi(0)} a \,, \qquad \text{and} \qquad \hb_{\sPi} = \tfrac{1}{\psi(0)} b\,.
\end{equation}
Thus $a\circ b$ is not identical with $a\cdot b$, but
\begin{equation}
  a \circ b = \tfrac{1}{\psi(0)} ab\,.
\end{equation}
Also we can show that $\circ$-product satisfies Leibniz rule
\begin{equation}
  \frac{\partial}{\partial X^I} (f\circ g) = \frac{\partial f}{\partial X^I} \circ g + f \circ \frac{\partial g}{\partial X^I} 
\end{equation}

In general, (\ref{level_matching_star_product}) can be generalized to arbitrary number of weakly constrained fields 
\begin{equation}
  \partial_{I} \partial^{I} \big(f_{1} \circ f_{2} \circ \cdots \circ f_{n}\big) = 0\,,
\end{equation}
and 
\begin{equation}
  f_1 \circ \cdots \circ \partial_{I} f_{i} \circ \cdots \circ \partial^I f_j \circ \cdots \circ f_n = 0\,, \qquad 1\leq i\leq j\leq n\,.
\label{strongConstraintLike}\end{equation}
The second property is reminiscent of the strong constraint, which is imposed by hand in strongly constrained DFT. On the other hand, (\ref{strongConstraintLike}) is an identity, thus the strong constraint is no longer necessary.

%%%%%%%%%%%%%%%%%%%%%%%%%%%%%%%%%
\subsection{Relation to the Hull-Zwibach's porjector}\label{sec:HZproj}
In \cite{Hull:2009mi}, Hull and Zwiebach introduced a projector in order to satisfy the level matching constraint for unconstrained double fields. It is defined by inserting an operator $\delta_{L_0 - \bar{L}_0,0}$ within a Fourier expansion. For massless fields, the $\delta_{L_0-\bar{L}_0,0}$ is represented by
\begin{equation}
    \delta_{L_0 - \bar{L}_0,0} = \delta_{\partial_I \partial^I,0}\,.
\end{equation}
The projector for an arbitrary unconstrained field $f$ is defined as
\begin{equation}
  \proj{f} = \sum_{K^I\in \mathbb{Z}^{2d}} \delta_{K_I K^I,0} \tilde{f}_K e^{2\pi i K_I X^I}\,,
\label{projection}\end{equation}
which is coming from string product in closed string field theory. It is obvious that the projector satisfies level matching constraint due to the Kronecker delta
\begin{equation}
  \partial_I \partial^I \proj{f} = 0\,.
\end{equation}
The projector for the usual product of two constrained fields $f$ and $g$ is given by 
\begin{equation}
  \proj{f \cdot g} = \sum_{K^I, K'^I} \delta_{K_I K'^I,0} \tilde{f}_K \tilde{g}_{K'} e^{2\pi i (K+K')_I X^I}\,,
\label{twoproj}\end{equation}  
where $K_I$ and $K'_I$ are null vectors. 
One can show that within the projector strong constraint is automatically satisfied
\begin{displaymath}
  \proj{\partial_I f \cdot \partial^I g} =0
\end{displaymath}
and it is commutative 
\begin{displaymath}
  \proj{f g} = \proj{g f}\,.
\end{displaymath}
However, the projector is not associative
\begin{equation}
  \proj{\proj{f g} h} \neq \proj{\proj{g h} f} \neq \proj{\proj{h f} g} \neq \proj{f g h}
\,.
\end{equation}

We can rewrite the projector of two weakly constrained fields (\ref{twoproj}) by using an inverse X-ray transform instead of the Fourier expansion
\begin{equation}
\begin{aligned}
  \proj{f \cdot g} & = \sum_{\Pi,\Pi'\in\cPo} \varphi(\Pi) \varphi(\Pi')\, \delta_{\partial_I \partial^I, 0} \,\hf_{\sPi}(z^i) \hg_{\sPi'}(z')
  \\
  & = \sum_{\Pi,\Pi'\in\cPo} \varphi(\Pi)\varphi(\Pi')\sum_{\ell,\ell'} \delta_{\ell_i \Pi^i{}_I \ell'_j \Pi'^{j I}, 0} \,\tilde{\hf}_{\sPi,\ell}\, \tilde{\hg}_{\sPi',\ell'} \, e^{2\pi i (\ell_i \Pi^i{}_I +\ell'_j \Pi'^{j}{}_I)X^I }\,,
\end{aligned}
\end{equation}
where $\tilde{\hf}_{\sPi,\ell}$ and $\tilde{\hg}_{\sPi',\ell'}$ are the Fourier modes on $d$-dimensional null-plane. In order to make sense the Kronecker-delta we impose a vanishing condition 
\begin{equation}
  \ell_i \ell'_j \,\Pi^i{}_I  \Pi'^{j I} = 0\,.
\label{lpilpi}\end{equation}
If $\Pi$ and $\Pi'$ are orthogonal, this condition is satisfied trivially. Nevertheless $\Pi$ and $\Pi'$ are not orthogonal, it is possible to satisfy (\ref{lpilpi}) due to Fourier zero-modes. As an example let us consider $O(2,2)$ case. If we assume that the $\hf_{\sPi}$ is depend only on $z^2$, $\hf_{\sPi}(z^2)$, and $\hg_{\sPi'}$ is depend only on $z'^1$, $\hg_{\sPi'}(z'^1)$, then the $\ell_2$ and $\ell'_1$ are remained and $\ell_1=\ell'_2 = 0$. If we denote $t^{ij} = \Pi^i{}_{I} \Pi'^{jI}$ and fix $t^{21} =0$, then the  $\ell_2 t^{21} \ell'_1$ vanish. The other elements also vanish due to the zero-modes
\begin{equation}
  \ell_1 t^{11} \ell'_{1} = \ell_1 t^{12} \ell'_2 = \ell_{2}t^{22} \ell'_{2} = 0\,.
\label{}\end{equation}
Hence, the zero mode contribution is missing in $\circ$-product. 

Therefore, we can separate HZ projector, $\proj{f\cdot g}$, into the associative part and the non-associative part as
\begin{equation}
  \proj{f\cdot g} = f\circ g + f \star g \,,
\end{equation}
where the $f \star g$ stands for the zero mode contribution. Of course, $f \star g$ satisfies the level matching constraint as well as $\partial^I f \star \partial_I g = 0$ , but it is not associative. In this sense, $\circ$-product is an associative truncation of the projector, even though it is not clear what truncation suppresses the zero-mode contribution.

%%%%%%%%%%%%%%%%%%%%%%%%%%%%%%%%%%%%%%%%%%%%%%%%
\section{$\OddZ$ transform and Polarization}
\subsection{$\Odd$ transformation}
Before discussing $\OddZ$ transform for weakly constrained fields, we will describe how $\OddZ$ group equipped with $\circ$-product is realized as a group action. To distinguish with the usual $\OddZ$ group, we denote it as $\OddZ_\circ$.

Let us introduce the $\OddZ_\circ$ metric $\cJ_\circ$ which is defined as
\begin{equation}
  \cJ_\circ = \mtr{0 & I_d \\ I_d & 0}\,,
\end{equation}
where the $d\times d$ identity matrix $I_d$ is defined by
\begin{equation}
  I_{d} = \sum_{\Pi} \varphi(\Pi) ~\mathbf{1}_{d}\,,
\end{equation}
where $\mathbf{1}_d = \text{diag}(1,\cdots,1)$. Note that $\cJ_{\circ}$ is a constant matrix, and it is different from the usual $\Odd$ metric
\begin{equation}
  \cJ_\circ{}_{IJ} \neq \mtr{0 & \delta^i{}_j \\ \delta_i{}^j & 0}\,.
\end{equation}
Now we define $\OddZ_\circ$ as the set of $2d\times 2d$ matrices satisfying 
\begin{equation}
  \cO^t \circ \cJ_\circ \circ \cO = \cJ _\circ \,,
\label{defOdd}\end{equation}
where $\cO \in \OddZ_\circ$.  Since $\cJ_\circ$ is also expanded by the inverse X-ray transform 
\begin{equation}
  \cJ_\circ = \sum_{\Pi} \varphi(\Pi) \hcJ_{\scriptscriptstyle\Pi} \,,
\end{equation}
where the $\hcJ_{\circ\sPi}$ is a usual $\OddZ$ metric which is independent with the $\Pi$
\begin{equation}
  \hcJ_{\circ\sPi} = \mtr{0 & \mathbf{1}_d \\ \mathbf{1}_d & 0}\,.
\label{JXray}\end{equation} 
As before, the $\cO$ may be reconstructed by an  inverse X-ray transform in terms of its X-ray images $\hcO_{\sPi}$
\begin{equation}
  \cO = \sum_{\Pi} \varphi(\Pi) \, \hat{\cO}_{\sPi}(z_i)\,,
\label{OddXray}\end{equation}
If we substitute the expansions (\ref{JXray}) and (\ref{OddXray}) into the definition of $\OddZ_\circ$ in (\ref{defOdd}), then we have
\begin{equation}
  \sum_{\Pi} \varphi(\Pi) \,(\hcO_{\sPi})^t \,\hcJ_{\sPi}\, \hcO_{\scriptscriptstyle \Pi} = \sum_{\Pi} \varphi(\Pi) \hcJ_{\sPi}\,.
\end{equation}
This implies that each X-ray image $\hcO_{\sPi}$ is an $\OddZ$ element
\begin{equation}
  \hcO_{\sPi}^t \cdot \hcJ_{\sPi} \cdot \hcO_{\sPi} = \hcJ_{\sPi}\,, 
\end{equation}
thus any $\OddZ_\circ$ element $\cO$ is represented by a sum of usual $\OddZ$ elements $\hcO_{\sPi}$. 
In the case of a constant $\cO$, the $\hcO_{\sPi}$ is independent of the $\Pi$ and is proportional to $\cO$ using (\ref{Xrayconstant}),
\begin{equation}
 \hcO_{\sPi} = \tfrac{1}{\psi(0)} \cO \,.
\end{equation}

We can now show that $\OddZ$ with a group operation $\circ$-product forms a group.
For arbitrary elements $\cO_1, \cO_2, \cO_3 \in \OddZ_\circ$, they satisfy the following the properties:
\begin{itemize}
  \item Closure
\begin{equation}
\cO_1\circ \cO_2 \in \OddZ_\circ  
\end{equation}
  \item Associativity
\begin{equation}
  \cO_1\circ(\cO_2\circ \cO_3) = (\cO_1\circ \cO_2)\circ \cO_3
\end{equation}
  \item Identity
\begin{equation}
  A \circ I_{2d} = I_{2d}\circ A = A
\end{equation}
  \item Inverse
\begin{equation}
  A \circ \inv{A} = \inv{A} \circ A = I_{2d}
\end{equation}
\end{itemize}
The closure and associativity can be easily shown by definition of $\circ$-product. 
It is also obvious the $I_{2d}$ is the identity matrix for $\OddZ_{\circ}$ group
\begin{equation}
  \cO \circ I_{2d} = I_{2d} \circ \cO = \sum_{\Pi} \varphi(\Pi)\, \mathbf{1}_d \cdot \hat\cO_{\sPi} = \cO
\end{equation}
The inverse element is defined as $\cO \circ \inv{\cO} = I_{2d}$, and it is expanded using  the inverse X-ray transform 
\begin{equation}
\sum_{\Pi^i} \varphi(\Pi) ~\widehat\cO_{\sPi}\, \widehat{\inv{\cO_{\sPi}}} = \sum_{\Pi^i} \varphi(\Pi^i) ~\mathbf{1}_{2d}
\end{equation}
This implies that the hat operator defined in (\ref{hatOp})  commutes with the inverse operation
\begin{equation}
  \widehat{\inv{\cO_{\sPi}}} = \inv{\big(\widehat{\cO}_{\sPi}\big)}\,,
\end{equation}
thus the inverse of an $\OddZ_\circ$ element is expanded by the inverse of $\OddZ$ elements
\begin{equation}
  \inv{\cO} = \sum_{\Pi} \varphi(\Pi) \, \inv{\hcO}_{\sPi}\,.
\end{equation}
Since there is inverse for each element of usual $\OddZ$, for $\cO \in \OddZ_\circ$, we can always define inverse $\inv{\cO}$.

Now we define an arbitrary rank $\OddZ_{\circ}$ tensors $T_{I_1 I_2\cdots I_n}$ which transforms under the $\Odd_\circ$ as 
\begin{equation}
  T'{}_{I_1 \cdots I_m}{}^{J_1 \cdots J_n} (X') = \cO_{I_1}{}^{K_1}\circ \cdots \circ\cO_{I_1}{}^{K_m} \circ T_{K_1 \cdots K_m}{}^{L_1 \cdots L_n} \circ \cO^{J_1}{}_{L_1}\circ \cdots \circ \cO^{J_n}{}_{L_n}\,.
\label{OddZ}\end{equation}
Here the $\OddZ_\circ$ vector indices $I,J,\cdots$ are raised and lowered by $\cJ_{\circ}$ analogous to strongly constrained DFT.

%%%%%%%%%%%%%%%%%%%%%%%%
\subsection{Physical degrees of freedom and polarization}\label{sec:polarization}
Let us consider how we might realize the physical degrees of freedom in a manifestly $\OddZ_\circ$ covariant way. As we have shown in the previous section, a weakly constrained field $f(X^I)$ is represented by summing over all possible X-ray images $\hf_{\sPi}(z^i)$, which are strongly constrained fields defined on closed $d$-dimensional null planes $\cD^0(X^I,\Pi)$. Conversely, one may consider a summation over all possible strongly constrained generalized metrics and dilatons
\begin{equation}
  \cH_{IJ}(X^I) = \sum_{\Pi\in \cP^0_d} \varphi(\Pi) \hcH_{\sPi}{}_{IJ}(z^i)\,, \qquad d(X^I) = \sum_{\Pi\in \cP^0_d} \varphi(\Pi) d_{\sPi}{}_{IJ}(z^i)\,,
\label{expansionHd}\end{equation}
We define the $\cH_{IJ}$ and $d$ as the weakly constrained generalized metric and dilaton respectively. One can show that the $\cH_{IJ}$ satisfies following conditions:
\begin{equation}
\cH_{IJ} = \cH_{(IJ)}\,, \qquad\text{and}\qquad \cH_{IJ} \circ \cJ^{JK} \circ \cH_{KL} = \cJ_{IL}\,.
\label{genMetricOdd}\end{equation}
As strongly constrained DFT one can solve these conditions if we assume that the upper left component is non-degenerate. The $\cH$ is parametrized in terms of weakly constrained component fields $g$ and $B$
\begin{equation}
  \cH = \mtr{g^{-1} & g^{-1} \circ B \\ B\circ g^{-1} & g- B\circ g^{-1} \circ B}\,,
\label{paraGenMetric}\end{equation}
where the $g^{-1}$ is defined by
\begin{equation}
  g^{-1} \circ g = g \circ g^{-1} = I_d\,.
\end{equation}

Now we consider the parametrization of dilaton $d$. The exponentiation of the $d$, $[e^{-2d}]_\circ$, is defined by 
\begin{equation}
\begin{aligned}
  \,[e^{-2d}]_\circ & := I - 2d + \half (2d) \circ (2d) - \tfrac{1}{3!} (2d) \circ (2d) \circ (2d) + \cdots
\\
& = \sum_{\Pi\in \cPo}\sum_{m\geq0} \varphi(\Pi)  \tfrac{1}{m!} \big(-2 \hat{d}_{\sPi}(z^i)\big)^m
\\
&= \sum_{\Pi\in\cPo} \varphi(\Pi) e^{-2\hd_{\sPi}} 	\,.
\end{aligned}
\end{equation}
As strongly constrained DFT, we want to express $[e^{-2d}]_\circ$ using $g_{ij}$ and $\phi$ as
\begin{equation}
  [e^{-2d}]_\circ = \sqrt{|g|} \circ [e^{-2\phi}]_\circ\,.
\end{equation}
Here $\sqrt{|g|}$ is defined by the inverse X-ray transform
\begin{equation}
  \sqrt{|g|} = \sum_{\Pi\in\cPo} \varphi(\Pi) \sqrt{|\hg_{\sPi}|} \,.
\end{equation}
Therefore, the physical degrees of freedom for weakly constrained DFT are same as strongly constrained DFT, but they depend on both momentum and winding coordinates
\begin{equation}
      g(x,\tx)\,,\qquad B(x,\tx)\,, \qquad \phi(x,\tx)\,,
\end{equation}
and each component fields satisfy the level matching constraint. This feature is consistent with the string field theory. However, it is not clear what the geometric meaning of the weakly constrained component fields is. For instance, $g(x,\tx)$ cannot be interpreted as usual metric. 

Now consider a parametrization of the X-ray images $\hcH_{\sPi}$ and $\hd_{\sPi}$ in (\ref{expansionHd}). The question arises naturally of how to impose a section condition for each slicing matrix $\Pi$. The $T^{2d}$ consists of a physical torus $T^d$ and a dual torus $\tilde{T}^d$, and each of these two tori are represented by maximal null subspaces. The section condition seperates the doubled torus into physical torus and its dual torus and ignores the dual torus dependence. A polarization $\Theta$ provides a consistent way to separate the $T^{d}$ and $\tT^{d}$ within the double torus $T^{2d}$\cite{HullSigma1,HullSigma2}. Thus parametrization of the $\hcH_{\sPi}$ and $\hd_{\sPi}$ using usual physical variables requires explicit polarization for each $\Pi$. 

In strongly constrained DFT, the fields depend only on the coordinate $z^i$ of the physical torus $T^d$. Since the X-ray image fields are functions of $z^i = \Pi^i{}_I X^I$, we can regard the null subspace $D^0(X^I,\Pi)$ as a physical torus $T^d$ with a coordinate $z^i$. Also, we introduce a dual coordinate $\tz^i$ corresponding to the dual torus $\tilde{T}^d$. Then the slicing matrix $\Pi$ defines polarization $\Theta$
\begin{equation}
  \Theta_{\hI}{}^{I} = \mtr{\Pi^{iI}\\ \tPi_{i}{}^{I}} \,,
\end{equation}
where $\tPi$ is a $d\times 2d$ matrix of rank $d$. The polarization $\Theta$ is an $\OddZ$ element satisfying
\begin{equation}
  \Theta_{\hI}{}^{I} \hcJ_{IJ} \big(\Theta^t\big)^{J}{}_{\hJ} = \hcJ_{\hI\hJ}\,.
\label{polariationOdd}\end{equation}
The doubled coordinate $X^I$ is also decomposed into the physical torus part $z^i$ and its dual torus part $\tilde{z}_i$ 
\begin{equation}
  X^{\hI} = \Theta^{\hI}{}_I X^I = \mtr{\tz_i\\ z^i}
\end{equation}

From (\ref{polariationOdd}) we can determine the $\tPi$
\begin{equation}
  \hcJ_{\hI\hJ} = \mtr{\Pi^{i I} \hcJ_{IJ}\Pi^{t}{}^{Jj} &~ \Pi^{i I} \hcJ_{IJ}\tPi^{t}{}^{J}{}_j\\ \tPi_{i}{}^{I} \hcJ_{IJ}\Pi^{t}{}^{Jj} &~ \tPi_{i}{}^{I} \hcJ_{IJ}\tPi^{t}{}^{J}{}_j} = \mtr{0 & \delta^i{}_j \\ \delta_i{}^j & 0}\,.
\end{equation}
The upper-left part is guaranteed by null property of $\Pi$. The upper-right (or lower-left) corner implies that $\tilde\Pi$ is a right-inverse of $\Pi$
\begin{equation}
  \Pi \, \tPi= I_d\,, \qquad \text{or} \qquad (\tPi^t)^{I}{}_i= (\inv{\Pi})^I{}_{i}
\end{equation}

The X-ray image $\hcH_{\sPi}{}_{IJ}$, which lives on a null-plane $\cD^{0}(X; \Pi)$, is parametrized by using the polarization $\Theta$ in terms of metric and Kalb-Ramond fields 
\begin{equation}
  \Theta_{\hI}{}^{I}\hcHp{}_{IJ} (\Theta^t)^{J}{}_{\hJ} = \hcH_{\sPi}{}_{\hI\hJ} =
 \mtr{\check{g}_{\sPi}{}^{ij} &~ -\check{g}_{\sPi}{}^{ik} \check{B}_{\sPi}{}_{kj} \\ \check{B}_{\sPi}{}_{ik} \check{g}_{\sPi}{}^{kj} &~ \check{g}_{\sPi}{}_{ij} - \check{B}_{\sPi}{}_{ik} \check{g}_{\sPi}{}^{kl}\check{B}_{\sPi}{}_{lj}}\,.
\label{polarizationcH}\end{equation}
This is the usual parametrization of the generaized metric in strongly constrained DFT. However, it is important to note that the $\hg_{\sPi}{}_{ij}$ is different to the $\check{g}_{\sPi}{}_{ij}$ and related by field redefinition. Hence, we have two kinds of polarizations. The first polarization is for the parametrization of weakly constrained generalized metric $\cH$ (\ref{paraGenMetric}). In this case, polarization is independent to the $\Pi$. The second polarization is for the parametrization of the X-ray images of $\cH$ in (\ref{polarizationcH}), which is defined on each null-planes $\cD^{0}(X;\Pi)$. These two polarizations are related by an $\OddZ$ rotation by $\Theta$ and we should distinguish between $\hg_{\sPi}{}_{ij}$ and $\check{g}_{\sPi}{}_{ij}$.

%%%%%%%%%%%%%%%%%%%%%%%
\subsection{$\OddZ_\circ$ transform of the physical fields}
Now let us consider $\OddZ_\circ$ transform of the physical degrees of freedom. Using the $\OddZ$ transform of the strongly constrained fields, $\hcH_{\sPi}$ and $\hd_{\sPi}$, we can show that the $\cH$ is a rank 2 tensor and $d$ is a scalar under the $\OddZ_\circ$ transform
\begin{equation}
  \cH \longrightarrow \cO \circ\cH \circ \cO^{t} \,, \qquad d \longrightarrow d\,.
\label{Oddcirc}\end{equation}
The $\OddZ_\circ$ transform is global rotation as strongly constrained DFT, and the $\cO$ is a constant element. 

As we have discussed in \ref{sec:InverseXray}, all the X-ray images of a constant field are equal regardless of the $\Pi$. Thus the $\cO$ is expanded as
\begin{equation}
  \cO = \sum_{\Pi\in\cPo} \varphi(\Pi)\, \hcO_{\sPi}\,,
\label{}\end{equation}
where 
\begin{equation}
  \hcO_{\sPi} := \hcO = \tfrac{1}{\psi(0)} \cO
\label{}\end{equation}

Then the $\OddZ$ transform (\ref{Oddcirc}) is rewritten as
\begin{equation}
  \hcH'_{\sPi} = \hcO \,\hcH_{\sPi} \,\hcO^t \,,
\label{}\end{equation}
and all the $\hcH_{\sPi}$ are rotated an equal amount. For example, when $\cO = \cJ_\circ$, which is $\OddZ_\circ$ metric, then we have
\begin{equation}
  \hcH'_{\sPi}{}_{IJ} = \hcJ_{IK} \, \hcH_{\sPi}{}^{KL} \hcJ_{LJ}\,,
\label{}\end{equation}
or after using the polarization $\Theta_{\hI}{}^{I}$
\begin{equation}
  \hcH'_{\sPi}{}_{\hI\hJ} = \hcJ_{\hI\hK} \, \hcH_{\sPi}{}^{\hK\hL} \hcJ_{\hL\hJ}
\end{equation}
Thus the Buscher rule for a weakly constrained field is obtained by summing over all the Buscher rule for the strongly constrained fields.

%%%%%%%%%%%%%%%%%%%%%%%%%%%%%%%%%%%%%%%%%%%%%%%%%%%
\section{Action and Gauge transform}
%%%%%%%%%%%%%%%%%%%%%%%
\subsection{Gauge transform}
We start by using the previous formulation, which is manifestly compatible with level matching constraint, to construct a consistent gauge transformation and corresponding gauge invariant action without using the strong constraint. As we have shown, the weakly constrained generalized metric $\cH$ and dilaton $d$ are reconstructed by their X-ray images. The X-ray image fields are strongly constrained generalized metric and dilaton on a null-plane. This suggests that the gauge transform of the $\cH$ and $d$ can be represented by adding the gauge transformations for strongly constrained DFT. In strongly constrained DFT, the gauge transform is given by generalized Lie derivative
\begin{equation}
\begin{aligned}
	\hcL_{\hX} \hcH_{IJ} & = \hX^K \partial_K \hcH_{IJ} + (\partial_{I} \hX^K - \partial^{K} \hX_I ) \hcH_{K J} + (\partial_{J} \hX^K - \partial^{K} \hX_J ) \hcH_{IK} 
	\\
	\hcL_{\hX} \hd & = \hX^K \partial_K \hd - \half \partial_I \hX^I
\end{aligned}\label{g.Lie.deriv}
\end{equation}
where the $\hX$ is a strongly constrained gauge parameter, and the $\hcH$ and $\hd$ are strongly constrained generalized metric and dilaton. 
The gauge algebra of the generalized Lie derivative is closed under the section condition or strong constraint
\begin{equation}
  \comm{\hcL_X}{\hcL_Y}\cH_{IJ} = \hcL_{\comm{X}{Y}_C} \cH_{IJ} - F_{IJ}
\end{equation}
where the $F_IJ$ is the terms which vanishes under the strong constraint
\begin{equation}
  F_{IJ} = \half X^K \partial_L Y_K \partial^L \cH_{IJ} + \partial_K X^L \partial^K Y_J \cH_{I L} + \partial_{K} X^L \partial^K Y_I \cH_{LJ} - (X\leftrightarrow Y)
\end{equation}
Thus the strong constraint is essential for closure of the gauge algebra.

Now we consider the gauge transform of the $\cH$ and $d$. As the $\cH$ and $d$ are weakly constrained, their gauge transform $\delta \cH$ and $\delta d$ must be weakly constrained as well 
\begin{equation}
 \partial_I \partial^I \,\delta \cH = 0\,, \qquad\text{and}\qquad \partial_I \partial^I \,\delta d = 0
\end{equation}
Moreover, the $\cH$ and $d$ are represented by a sum of all possible strongly constrained generalized metrics $\hcH_{\sPi}$ and dilaton $\hd_{\sPi}$ respectively. Hence the gauge transform should be represented as a sum of generalized Lie derivatives. 

Using the properties of $\circ$-product, one can speculate that infinitesimal gauge transformation of the $\cH$ and $d$ takes the form
\begin{equation}
\begin{aligned}
    \delta_X \cH_{IJ} & = X^{K} \circ \partial_{K}\cH_{IJ} + \big(\partial_{I} X^{K} - \partial^{K} X_{I}\big) \circ \cH_{KJ} + \big(\partial_{J} X^{K} - \partial^{K} X_{J}\big) \circ \cH_{IK}\,,
\\
\delta_X d & = X^{I} \circ \partial_{I} d - \half \partial_{I} X^{I}\,,
\end{aligned}\label{gaugetransf}
\end{equation}
where the $X^I$ is a weakly constrained gauge parameter. This gauge transformation is  analogous to generalized Lie derivative in (\ref{g.Lie.deriv}), but the ordinary product is replaced by $\circ$-product. Also, one can show that the $[e^{-2d}]_\circ$ transform like a density 
\begin{equation}
  \delta_X [e^{-2d}]_\circ = X^{I} \circ \partial_{I} [e^{-2d}]_\circ + \partial_I X^I \circ [e^{-2d}]_\circ\,.
\end{equation}
We can easily generalize the gauge transformations to the arbitrary rank of $\OddZ_\circ$ tensor $T_{I_1 I_2 \cdots I_n}$ as
\begin{equation}
  \delta_X T_{I_1 I_2 \cdots I_n} = X^{J} \circ \partial_{J} T_{I_1 I_2 \cdots I_n} + \sum_{i=1}^{n} \big(\partial_{I_i}X^J - \partial^J X_{I_i}\big) \circ T_{I_1 \cdots I_{i-1} J I_{i+1} \cdots I_{n}}
\end{equation}
A priori this gauge transform does not move the X-ray image fields $\hT_{\sPi}$ defined on a null-plane $\cD^0(X;\Pi)$ to another field defined on a different null-plane $\cD^0(X;\Pi')$, where $\Pi \neq \Pi'$. In other words, the gauge transform preserves the section condition for a given set of X-ray image fields
\begin{equation}
  \hT_{\sPi} \longrightarrow \hT'_{\sPi}
\label{}\end{equation}

It is straightforward to show that the algebra of the gauge transformations (\ref{gaugetransf}) is closed exactly without using strong constraint
\begin{equation}
  \comm{\delta_X}{\delta_Y}\cH_{MN} = \delta_{\comm{X}{Y}_{C_\circ}} \cH_{MN}\,,
\end{equation}
where the $C_\circ$-bracket is defined as
\begin{equation}
  \comm{X}{Y}^M_{C_\circ}=  X^{N}\circ \partial_N Y^M - \half X^N \circ \partial^M Y_{N} - \big( X \leftrightarrow Y \big)\,.
\end{equation}
This can be easily proved by noting that the $\circ$-product satisfies the identity (\ref{strongConstraintLike}), which is similar with the strong constraint, and other algebraic properties of $\circ$-product (\ref{algebraicProperties}). 

One can rewrite the gauge transform using the explicit parametrization of the $\cH$ in (\ref{paraGenMetric}) and the weakly constrained $\OddZ_\circ$ covariant gauge parameter 
\begin{equation}
  X^I = \mtr{\Lambda_i \\ \xi^i}\,.
\end{equation}
After the $\tpartial$-expansion, the gauge transform  takes the form
\begin{equation}
\begin{aligned}
  \delta^{\order{0}} \cE_{ij} &= \partial_i \Lambda_j - \partial_j \Lambda_i + \xi^k \circ \partial_k \cE_{ij} + \partial_i \xi^k \circ\cE_{kj} + \partial_j \xi^k \circ\cE_{ik}\,,
  \\
  \delta^{\order{1}} \cE_{ij} &= - \cE_{ik}\circ \big(\tilde{\partial}^k \xi^l - \partial^{l}\xi^{k} \big)\circ \cE_{lj} + \Lambda_k\circ \tpartial^k \cE_{ij} - \tpartial^k \Lambda_i \circ\cE_{kj} - \tpartial^k \Lambda_j \circ\cE_{ik} \,,
\end{aligned}\label{gaugetransformE}
\end{equation}
where $\cE = g + B$. This result is the similar to Hohm, Hull and Zwiebach's tilde-derivative expansion \cite{Hohm:2010jy} except the $\circ$-product. Unlike strongly constrained DFT, we cannot identify the $\Lambda$ and $\xi$ as the 1-form gauge transform and diffeomorphism parameters, and the physical interpretation of the gauge transform is not obvious with this form. However, the inverse X-ray transform with the polarization suggests that the gauge transform of the $g$ and $B$ (\ref{gaugetransformE}) is just a sum of the usual diffeomorphisms and one-form gauge transforms over all possible null-planes $\cD^{0}(X;\Pi)$
\begin{equation}
\begin{aligned}
  \delta \check{g}_{\sPi} & = \cL_{\check{\xi}_{\sPi}} \check{g}_{\sPi}
  \\
  \delta \check{B}_{\sPi} & = 	\cL_{\check{\xi}_{\sPi}} \check{B}_{\sPi} + \tfrac{\partial \check\Lambda_j}{\partial z^{i}} - \tfrac{\partial \check\Lambda_i}{\partial z^{j}}
\end{aligned}\label{gaugetransfgB}
\end{equation}
where $\check{g}_{\sPi}$ and $\check{B}_{\sPi}$ are component fields of the $\hcH_{\sPi}{}_{\hI\hJ}$ and the $\cL_{\check{\xi}_{\sPi}}$ is the ordinary Lie derivative. Hence, the polarization provides a right basis of the physical degrees of freedom and simplifies the gauge transform for each X-ray image fields.

%%%%%%%%%%%%%%%%%%%%%%%
\subsection{Action}
We now construct an action which is invariant under the gauge transformations (\ref{gaugetransf}) and the $\OddZ_\circ$ transformations (\ref{Oddcirc}) in terms of weakly constrained fields $\cH$ and $d$. As we have seen, since the gauge transformation of $\cH$ and $d$ are represented by a sum of generalized Lie derivatives, we can easily speculate that the weakly constrained DFT action is also given by adding strongly constrained DFT actions defined on  null-planes. 
 
Before considering the action of weakly constrained DFT, let us recall the action of strongly constrained DFT. If we denote the strongly constrained generalized metric and dilaton as $\hcH$ and $\hd$ respectively, the strongly constrained DFT action is given by \cite{Hohm:2010pp}
\begin{equation}
  \cS_{\scriptscriptstyle\text{SDFT}} = \int_{T^{2d}} \dd^{2d}X e^{-2\hd} \,\cL_{\scriptscriptstyle\text{SDFT}}\,,
\end{equation}
where
\begin{equation}
\begin{aligned}
  \cL_{\scriptscriptstyle\text{SDFT}} = ~& 4 \hcH^{IJ}  \partial_{I} \partial_{J} \hd - \partial_{I} \partial_{J} \hcH^{IJ} - 4\hcH^{IJ} \partial_{I} \hd \partial_{J} \hd + 4\partial_{I} \hcH^{IJ} \partial_{J} \hd
\\
& + \tfrac{1}{8} \hcH^{IJ} \partial_{I}\hcH^{KL} \partial_{J} \hcH_{KL} - \half \hcH^{IJ} \partial_{I} \hcH^{KL} \partial_{K} \hcH_{JL}	
\end{aligned}\label{SDFTaction}
\end{equation}
However, the action and the associated gauge transformations are not enough to define a consistent theory. In strongly constrained DFT, section condition or strong constraint is essential for the consistency of the theory. It is an additional constraint which should be imposed by hand. Under the strong constraint all the interesting stringy information vanish, such as winding modes, and DFT reduces to the conventional supergravity at least locally.

We now propose a gauge invariant action for weakly constrained DFT as follows:
\begin{equation}
  \cS_{\scriptscriptstyle\text{WDFT}} = \int \dd^{2d} X\, \cL_{\scriptscriptstyle\text {WDFT}}
\label{actionWDFT}\end{equation}
where the Lagrangian $\cL_{\text{WDFT}}$ is given by
\begin{equation}
\begin{aligned}
  \cL_{\scriptscriptstyle\text {WDFT}} = ~&[e^{-2d}]_{\circ} \circ \Big[ 4 \cH^{IJ} \circ \partial_{I} \partial_{J} d - \partial_{I} \partial_{J} \cH^{IJ} - 4\cH^{IJ}\circ \partial_{I}d \circ \partial_{J} d + 4\partial_{I} \cH^{IJ}\circ \partial_{J} d
\\
& + \tfrac{1}{8} \cH^{IJ} \circ\partial_{I}\cH^{KL} \circ\partial_{J} \cH_{KL} - \half \cH^{IJ} \circ\partial_{I} \cH^{KL} \circ\partial_{K} \cH_{JL}	\Big]
\end{aligned}\label{WDFTaction}
\end{equation}
In this formula we have replaced the ordinary products to $\circ$-products in the strongly constrained DFT action (\ref{SDFTaction}). The $\cH$ and $d$ are given by weakly constrained fields, which are depend on both momenta and windings explicitly. Furthermore, the invariance of this action under the $\OddZ_\circ$ in (\ref{Oddcirc}) and the gauge transform in (\ref{gaugetransf}) is guaranteed by the identity \ref{strongConstraintLike}) and other useful algebraic properties. 

This Lagrangian itself is a weakly constrained field, thus it can be reconstructed by its X-ray images
\begin{equation}
  \cL_{\scriptscriptstyle \text{WDFT}} = \sum_{\Pi\in\cPo} \varphi(\Pi) \hcL_{\sPi}(z^i)\,,
\label{Xrayaction}\end{equation}
Using the definition of $\circ$-product, the X-ray images of the $\hcL_{\sPi}$ are given by
\begin{equation}
\begin{aligned}
  \hat{\cL}_{\sPi} & = e^{2 \hd_{\Pi}}\Big(  4 \hcH_{\sPi}{}^{IJ} \partial_I \partial_J \hd_{\sPi} - \partial_{I} \partial_{J} \hcH_{\sPi}{}^{IJ} - 4 \hcH_{\sPi}{}^{IJ} \partial_{I}\hd\partial_{J} \hd_{\sPi} 
\\
&\qquad + 4 \partial_{I} \hcH_{\sPi}{}^{IJ} \partial_{J} \hd + \tfrac{1}{8} \hcH_{\sPi}{}^{IJ} \partial_{I}\hcH_{\sPi}{}^{KL} \partial_{N} \hcH_{\sPi}{}_{KL} - \half \hcH_{\sPi}{}^{IJ} \partial_{I} \hcH_{\sPi}{}^{KL} \partial_{K} \hcH_{\sPi}{}_{JL}\Big)\,,
\end{aligned}\label{Spi}
\end{equation}
This form of the action is similar with the strongly constrained DFT action in (\ref{SDFTaction}), but it cannot be identified yet. In order to define a strongly constrained DFT section condition should be imposed, and the section condition requires the polarization. In section \ref{sec:polarization}, we have introduced the polarization for X-ray images of the weakly constrained generalized metric and dilaton, $\hcH_{\sPi}$ and $\hd_{\sPi}$ as
\begin{equation}
  \hcH_{\hI\hJ} =  \Theta_{\hI}{}^{I}\hcHp{}_{IJ} (\Theta^t)^{J}{}_{\hJ}\,.
\end{equation}
Then the previous Lagrangian (\ref{Spi}) is rewritten as 
\begin{equation}
\begin{aligned}
  \hat{\cL}_{\sPi} & = e^{2 \hd_{\Pi}}\Big(  4 \hcH_{\sPi}{}^{\hI\hJ} \partial_{\hI} \partial_{\hJ} \hd_{\sPi} - \partial_{\hI} \partial_{\hJ} \hcH_{\sPi}{}^{\hI\hJ} - 4 \hcH_{\sPi}^{IJ} \partial_{I}\hd\partial_{J} \hd_{\sPi} 
\\
&\quad + 4 \partial_{I} \hcH_{\sPi}{}^{IJ} \partial_{J} \hd + \tfrac{1}{8} \hcH_{\sPi}{}^{IJ} \partial_{I}\hcH_{\sPi}{}^{KL} \partial_{N} \hcH_{\sPi}{}_{KL} - \half \hcH_{\sPi}{}^{IJ} \partial_{I} \hcH_{\sPi}{}^{KL} \partial_{K} \hcH_{\sPi}{}_{JL}\Big)\,,
\end{aligned}
\end{equation}
where the $\hpartial_{\hI}$ is defined as
\begin{equation}
  \hpartial_{\hI} = \Theta_{\hI}{}^{I} \partial_I\,.
\end{equation}
The $\hcL_{\sPi}$ is a strongly constrained DFT Lagrangian defined on a null-plane $\cD^{0}(X^I,\Pi)$ with an oblique section condition 
\begin{equation}
  \frac{\partial}{\partial \tilde{z}_i} =0\,,
\end{equation}
where $\tilde{z}_i$ is a coordinate for dual torus $\tilde{T}$ which is separated by the polarization $\Theta^{\hI}{}_{I}$
\begin{equation}
  \tilde{z}_i = \tilde{\Pi}_{iI}X^I\,.
\end{equation}
Therefore, we propose that
\begin{quotation}
  \emph{Weakly constrained DFT is given by a sum of all possible strongly constrained DFT.}
\end{quotation}
One of the advantage of this formalism is that we can describe weakly constrained DFT in terms of what we already know. Since the weakly constrained DFT consists of the strongly constrained DFTs, we can easily extend all the strongly constrained DFT results to the weakly constrained DFT.

Let's consider the component field expression of the action. We apply the tilde-derivative expansion \cite{Hohm:2010jy} of the action as in the gauge transformation case. First, the zeroth-order is given by
\begin{equation}
\begin{aligned}
  \cL^{\order{0}} = [e^{2d}]_\circ \circ \Big[ &-\tfrac{1}{4} g^{ik}\circ g^{jl}\circ g^{pq} \circ \big( \partial_p \cE_{kl} \circ \partial_{q} \cE_{ij} - \partial_i \cE_{lp} \circ \partial_{j} \cE_{kq} - \partial_i \cE_{pl} \circ \partial_{j} \cE_{qk}\big)
  \\&
  + 2 \partial^i d \circ \partial^j g_{ij} + 4 \partial^i d \circ \partial_i d \Big]\,,
\end{aligned}\label{zerotilde}
\end{equation}
where $\partial^i = g^{ij} \circ\partial_j$. If we assume that all the fields are defined on $\tx=0$ plane or taking $\alpha' \rightarrow 0$ limit, then $\circ$-product reduces to the ordinary product. Then the previous action reduces to the string NSNS sector effective action
\begin{equation}
  \cL_{\rm NSNS} =\sqrt{g} e^{-2\phi} \Big( R - \tfrac{1}{12} H^2 + 4 \partial_i \phi \partial^i \phi\Big)
\label{}\end{equation}
The next order takes the form
\begin{equation}
\begin{aligned}
    \cL^{\order{1}} = [e^{2d}]_\circ \circ \Big[ & \half g^{ik}\circ g^{jl}\circ g^{pq}\circ \big(\cE_{pr} \circ \tpartial^{r} \cE_{kl} \circ \partial_q \cE_{ij} - \cE_{ir} \circ \tpartial^r \cE_{ip} \circ\partial_{k}\cE_{jq}
    \\&
    + \cE_{rl}\circ\tpartial^r \cE_{pi} \circ \partial_{k}\cE_{qj}\big) + g^{ip} \circ g^{jq}\circ \big(\cE_{rq} \circ \partial_p d \circ \tpartial^r \cE_{ij} - \cE_{pr} \circ \tpartial^r d\circ \partial_q \cE_{ij} 
    \\&
    + \cE_{rp} \circ\tpartial^{r} d \circ\partial_{q} \cE_{ij} - \cE_{qr}\circ \partial_{p} d \circ\tpartial^r \cE_{ji} 
    \big) - 8 g^{ij} \circ \cE_{ik} \circ \tpartial^k d \circ \partial_j d
     \Big]\,,
\end{aligned}\label{onetilde}
\end{equation}
and finally the $\tilde{\partial}^2$ order is given by
\begin{equation}
\begin{aligned}
    \cL^{\order{2}}  = [e^{-2d}]_{\circ} \circ \Big[&-\tfrac{1}{4} g^{ik}\circ g^{jl}\circ g^{pq} \circ \big( \cE_{pr}\circ \cE_{qs}\circ \tpartial^r \cE_{kl} \circ \tpartial^{s} \cE_{ij} - \cE_{ir} \circ\cE_{js}\circ \tpartial_r \cE_{lp} \circ \tpartial^{s} \cE_{kq} 
    \\&
    - \cE_{ri}\circ\cE_{sj} \circ \tpartial^r \cE_{pl} \circ \tpartial^{s} \cE_{qk}\big) + 4 g^{ij}\circ \cE_{ik} \circ\cE_{jl} \circ\tpartial^k d\circ \tpartial^l d
    \\&
    - g^{ik}\circ g^{jl} \circ \big( \cE_{ip}\circ \cE_{qj} \circ\tpartial^{p} d \circ\tpartial^{q}\cE_{kl} + \cE_{pi}\circ \cE_{jq} \circ \tpartial^{p} d \circ\tpartial^{q} \cE_{lk} \big)
    \Big]
\end{aligned}\label{twotilde}
\end{equation}
This result is consistent with \cite{Hohm:2010jy}. However, the physical interpretation is not apparent with this form.

Analogous to the gauge transform, we can rewrite these Lagrangians by inverse X-ray transform. Using the strongly constrained component fields $\check{g}_{\sPi}$, $\check{B}_{\sPi}$ and $\check{\phi}_{\sPi}$ after imposing the polarization, the Lagrangians reduce reduce to
\begin{equation}
  \cL = \sum_{\sPi\in\cPo} \varphi(\Pi) \hcL_{\sPi}\,,
\end{equation}
where the X-ray image of the Lagrangian is given by
\begin{equation}
  \hcL_{\sPi} = \sqrt{\check{g}_{\sPi}} e^{\check{\phi}_{\scriptscriptstyle{\Pi}}} \Big(\check{R}_{\sPi} - \tfrac{1}{12} (\check{H}_{\sPi})^2 + 4 \check\partial_i \check{\phi}_{\sPi} \check\partial^i \check{\phi}_{\sPi} \Big)\,,
\end{equation}
where the $\check{R}_{\sPi}$ denotes the usual Ricci scalar and the $\check{\partial}_{i}$ denotes the partial derivative on a null-plane $\cD^{0}(X;\Pi)$ 
\begin{equation}
  \check{\partial}_{i} = \tfrac{\partial}{\partial z^i} = \Pi_i{}^{I} \tfrac{\partial}{\partial X^I}\,.
\end{equation}
Here the three-form field strength $\check{H}_{\sPi}$ is defined as
\begin{equation}
  \check{H}_{\sPi}{}_{ijk} = 3 \check{\partial}_{[i} \check{B}_{\sPi}{}_{jk]}\,.
\end{equation}
Therefore, under the field redefinition to the checked variables, the tilde-derivative expansion results \eqref{zerotilde}, \eqref{onetilde} and \eqref{twotilde} are now rewritten in the physically meaningful expression. 

Since weakly constrained DFT is represented by a sum of strongly constrained DFT, it is easy to see that a solution of equation of motion for weakly constrained DFT is also reconstructed by the strongly constrained DFT solutions.  For example, a collection of pp-wave solutions propagating to winding directions would be a well-defined solution of weakly constrained DFT. In \cite{NullWave1,NullWave2,NullWave3}, such pp-waves in doubled space are identified as fundamental strings. Hence by stacking the known pp-wave solutions in doubled torus we may construct a large class of string solutions for weakly constrained DFT.

%For consistency, weakly constrained DFT should reproduce the strongly constrained DFT by truncating the winding modes. If we take $\alpha' \rightarrow 0$ limit, the winding modes become very massive and negligible. In this limit, the only remaining degrees of freedom is momentum mode and the weakly constrained DFT reduces to the usual supergravity or strongly constrained DFT. To see this limit explicitly, we now restore the hidden $\alpha'$ factors by replacing the dimensionless coordinates to the dimensionful one. Then the doubled momentum $K=(k,\tk)$ are replaced as
%
%\begin{equation}
%  k \longrightarrow \frac{p}{R} \,, \qquad \tk \longrightarrow \frac{R}{\alpha'} w
%\label{}\end{equation}
%
%where the $p$ and $w$ are integers.  Hence, weakly constrained DFT reduces to strongly constrained DFT under the supergravity limit. 

%%%%%%%%%%%%%%%%%%%%%%%%%
\subsection{Example: $(1+1)$-dimensional case}
Here we exhibit an simplest example, the $d=1$ case. In the $\mathbf{O}(1,1)$ weakly constrained DFT, there is no Kalb-Ramond $B$-field, and the physical degrees of freedom are just the scalar metric $g$ and dilaton $\phi$. 

Weakly constrained fields $f(X^I)$ are written by adding one-dimensional X-ray images
\begin{equation}
  f(X^I) = \sum_{\Pi\in\cP^{0}_1} \varphi(\Pi)\hf_{\sPi}(z^i)
\end{equation}
Here $\cP^0_1$ is a set of $1\times 2$ matrices whose Smith normal forms are given by
\begin{equation}
  \Pi = U D_0 V 
\end{equation}
where $U$ is $\pm 1$ and $V\in \mathbf{O}(1,1;\mathbb{Z})$.
If we ignore signs, there are only two $\mathbf{O}(1,1;\mathbb{Z})$ elements
\begin{equation}
  V_1 = \mtr{1 ~& 0 \\ 0 ~& 1}\,, \qquad \text{and} \qquad V_2 = \mtr{0 ~& 1\\1 ~& 0}\,,
\end{equation}
thus there exists only two possible slicing matrices $\Pi$ 
\begin{equation}
  \Pi_1 = \mtr{ 1&~ 0}\,, \qquad \text{and} \qquad \Pi_2 = \mtr{0 &~ 1}\,.
\end{equation}
This means that in (1+1) dimensions there are only two X-ray image fields. Hence, any weakly constrained fields are reconstructed by two X-ray images
\begin{equation}
  f = \hf_{\sPi_1}(z_1) + \hf_{\sPi_2}(z_2)\,,
\label{}\end{equation}
where we have ignored the weight factor $\varphi$ since it is a finite sum in $\mathbf{O}(1,1)$ case. The coordinate $z_1$ and $z_2$ are represented
\begin{equation}
  z_1 = \Pi_1 X = x\,, \qquad z_2 = \Pi_2 X = \tx
\end{equation}

The right inverses for the slicing matrices are 
\begin{equation}
  \tPi_1 = \mtr{0 & ~1}\,, \qquad   \tPi_2 = \mtr{1 & ~0}\,.
\end{equation}
Then the corresponding polarizations $\Theta_{1,2}$ are written in terms of $\Pi_{1,2}$
\begin{equation}
  \Theta_1 = \mtr{\Pi_1 \\ \tPi_1 } = \mtr{1&0\\0&1} = \mathbf{1}_{2d} \,, \qquad \Theta_2 = \mtr{\Pi_2 \\ \tPi_2} = \mtr{0 &1\\ 1& 0} = \cJ\,.
\end{equation}
Hence the $\Theta_{1,2}$ are $\OddZ$ elements. 
The polarization of generalized metric is given by
\begin{equation}
\begin{aligned}
	\hcH_{1}{}_{\hI\hJ} = \Theta_1{}_{\hI}{}^{I} \hcH_{1}{}_{IJ} \Theta_1{}^{j}{}_{\hJ} = \hcH_{IJ}	
	\\
	\hcH_{2}{}_{\hI\hJ} = \Theta_2{}_{\hI}{}^{I} \hcH_{2}{}_{IJ} \Theta_2{}^{j}{}_{\hJ} = \cJ \hcH_{IJ}	\cJ
\end{aligned}
\end{equation}
where $\hcH_{1} = \hcH_{\sPi_1}$ and $\hcH_{2} = \hcH_{\sPi_2}$.
Since the X-ray images for $\cH$ are 
\begin{equation}
  \hcH_{1}{}_{IJ} = \mtr{\inv{\hg}_1 & 0 \\ 0 & \hg_1} \,, \qquad \hcH_{2}{}_{IJ} = \mtr{\inv{\hg}_2 & 0 \\ 0 & \hg_2}\,,
\end{equation}
the parametrization of the polarized generalized metrics $\hcH_{1,2}$ are as follows:
\begin{equation}
  \hcH_{1}{}_{\hI\hJ} = \mtr{\inv{\hg}_1 & 0 \\ 0 & \hg_1} = \mtr{\inv{\check{g}}_1 & 0\\0 & \check{g}_1} \,, \qquad \hcH_{2}{}_{\hI\hJ} = \mtr{\hg_2 & 0 \\ 0 & \inv{\hg}_2} = \mtr{\inv{\check{g}}_2 & 0\\0 & \check{g}_2} 
\end{equation}
thus we can identify $\check{g}_{1} = \hg_{1}$ and $\check{g}_{2} = \inv{\hg}_{2}$. Hence, the $\hcH_{1}$ is parametrized trivially, and the $\hcH_2$ is the $T$-dualized one. 

Lagrangian is also separated as follows: 
\begin{equation}
  \cL(x,\tx) = \hcL_1 (x) + \hcL_2 (\tx)\,.
\end{equation}
The $\hcL_1$ and $\hcL_2$ are usual strongly constrained DFT action with different section conditions, $\frac{\partial}{\partial \tx}=0$ and $\frac{\partial}{\partial x}=0$. Therefore, in $(1+1)$ dimensional case, there is no interaction between momentum and winding modes, and these two sectors are completely decoupled.

%%%%%%%%%%%%%%%%%%%%%%%%
\subsection{Relation to the closed string field theory}
Given the above discussion, a natural question is under what conditions do we expect the action to give a reasonable description of the massless degrees of freedom of string theory. In our construction, we have ignored massive string states with masses $m_s \simeq 1/\sqrt{\alpha'}$ but kept Kaluza-Klein momentum modes with $m_{KK} \simeq 1/R$ and string winding modes with $m_{w} \simeq R/(\alpha')^2$. Also, we have treated the momentum modes and winding modes on an equal footing. Thus the compactification scale should be of order of self-dual radius $R \simeq \sqrt{\alpha}$. Then the all the mass scales are of the same order, $m_s \simeq m_{KK} \simeq m_w$, and there is no specific limit which truncates massive string states only. 

According to the original derivation of DFT in \cite{Hull:2009mi}, weakly constrained DFT is assumed to be obtained by integrating out all the unwanted degrees of freedom, including massive string states, from closed string field theory. Since the mass scale of unwanted degrees of freedom is not heavier than the momentum and winding, the massive string states are not decoupled. Thus WDFT is not the usual Wilsonian effective theory, and effective field theory description is not guaranteed. However, our previous construction suggests that there exists a well-defined field theory for the weakly constrained DFT limit of closed string field theory.

Here, we want to compare fluctuations of our action (\ref{WDFTaction}) around constant backgrounds to the fluctuations of the closed string field theory. Up to cubic order fluctuations for closed string field theory action is computed in \cite{Hull:2009mi} by simply ignoring the string massive modes
\begin{equation}
\begin{aligned}
  \cS^{\order{3}}_{\rm \scriptscriptstyle CSFT}[e, d, \cdot\,] = \int \dd^{2d}X \Big[&\tfrac{1}{4} e_{ij} \cdot \Box e^{ij} + \tfrac{1}{4} (\bar{D}^j e_{ij})^2 + \tfrac{1}{4} (D^j e_{ij})^2 - 2 d \cdot D^{i} \bar{D}^j e_{ij} - 4d\cdot \Box d 
  \\
  &+ \tfrac{1}{4} e_{ij} \big(D^i e_{kl} \cdot \bar{D}^j e^{kl} - D^i e_{kl} \cdot \bar{D}^l e^{kj} - D^k e^{il} \cdot \bar{D}^j e_{kl}\big)
  \\
  &+\half d\cdot \big((D^i e_{ij})^2 + \bar{D}^j e_{ij})^2 + \half (D_{k} e_{ij})^2 + \half (\bar{D}_k e_{ij})^2 
  \\
  &\qquad\quad+ 2 e^{ij}\cdot (D_i D^k e_{kj} + \bar{D}_j \bar{D}^k e_{ik}) \big)
  \\
  & + 4 e_{ij} \cdot d \cdot D^i \bar{D}^j d + 4 d \cdot d \cdot \Box  d~
  \Big]\,,
\end{aligned}\label{flucCSFT}
\end{equation}
where the $e_{ij}(x,\tx)$ is the fluctuation field for metric and Kalb-Ramond field
\begin{equation}
  e_{ij} = h_{ij} + b_{ij}\,,
\end{equation}
and the $i,j,\cdots$ indices are raised and lowerd by a constant background metric $G_{ij}$. Here the derivative operators are defined as
\begin{equation}
\begin{aligned}
  & D_i = \partial_i - \tilde{\partial}_i - B_{ik} \tilde{\partial}^k\,,\qquad \bar{D}_i = \partial_i + \tilde{\partial}_i - B_{ik} \tilde{\partial}^k\,,
  \\
  &\Box = \partial^2 + \tilde\partial^2\,.
\end{aligned} 
\end{equation}
Interestingly, this action is gauge invariant even if massive states contribution is just ignored. Also, from higher than cubic order fluctuations HZ projector is explicitly involved, and it would be a non-associative theory. 

For weakly constrained DFT case, let us consider fluctuations of weakly constrained generalized metric and dilaton around constant backgrounds $\cH^0$ and $d^0$
\begin{equation}
  \cH_{IJ} = \cH^{0}_{IJ} + \bar{\cH}_{IJ}\,, \qquad d = d^{0} + \bar{d}\,.
\end{equation}
If we substitute these ansatz into (\ref{WDFTaction}) and keep up to cubic order terms, then we have the similar form with (\ref{flucCSFT})
\begin{equation}
  \cS^{\order{3}}_{\rm \scriptscriptstyle  WDFT} = \cS^{\order{3}}_{\rm \scriptscriptstyle  CSFT}[e, d, \circ\,] \,,
\label{flucWDFT}\end{equation}
where $\cS^{\order{3}}_{\rm \scriptscriptstyle CSFT}[e, d, \circ\,]$ means that the usual product in (\ref{flucCSFT}) is replaced by $\circ$-product. However, one can show that these two actions are not equivalent
\begin{equation}
  \cS^{\order{3}}_{\rm \scriptscriptstyle  CSFT}[e, d, \cdot \,] \neq \cS^{\order{3}}_{\rm \scriptscriptstyle  CSFT}[e, d, \circ\,]\,.
\label{}\end{equation}

For arbitrary weakly constrained fields $f,g,$ and $h$, we can show that 
\begin{equation}
    \int \dd^{2d}X ~ \tfrac{1}{\psi} f \cdot g = \int \dd^{2D}X ~f\circ g\,,
\end{equation}
however, from the product of three fields, the integration of $\circ$-products is not identical with the usual products
\begin{equation}
    \int \dd^{2d}X ~ \tfrac{1}{\psi^2} f \cdot g \cdot h \neq \int \dd^{2d} X ~f\circ g \circ h = \int \dd^{2d}X ~f\cdot (g \circ h)\,.
\end{equation}
Note that the Hull and Zwiebach's projector satisfy
\begin{equation}
  \int \dd^{2d}X f \cdot \proj{g h} = \int \dd^{2d}X ~f\cdot g \cdot h\,.
\end{equation}
This implies that the $\cS^{\order{3}}_{\rm \scriptscriptstyle CSFT}[e,d,\cdot]$ and $\cS^{\order{3}}_{\rm \scriptscriptstyle WDFT}$ are not equivalent. 

At this point it is not obvious what is the consistent perturbation of closed string field theory after integrating out all the massive string states. As already mentioned, it is practically impossible computation. From the cubic order perturbation, the mixing between string massless and massive modes arise. If we integrate out the massive mode fields, then $\cS^{\order{3}}_{\rm \scriptscriptstyle CSFT}[e,d,\cdot]$ would be modified. Therefore, it remains unclear whether the comparison to the closed string field theory computation makes sense. Now we wish to propose that at least our construction using $\circ$-product may provide a well-defined field theory description of massless subsector for closed string field theory. As we have discussed in section \ref{sec:HZproj}, the difference between HZ projector and $\circ$-product is given by zero-modes in Fourier expansion. If the Fourier zero-modes vanish under a certain truncation, then the projector reduces to $\circ$-product. In this sense, our formalism is an associative truncation of closed string field theory. However, it is not clear under what circumstances the zero-modes are suppressed, and this remains an open question.

%%%%%%%%%%%%%%%%%%%%%%%%%%%%%%%%%%%%%%%%%%%%%%%%%%%%%%%
\section{Conclusion}
In this paper we have shown that  the X-ray transform is well-suited for describing weakly constrained DFT. Any weakly constrained fields are represented by strongly constrained fields through the inverse X-ray transform. The key ingredient in our formalism is the $\circ$-product. The $\circ$-product is a binary operation for weakly constrained fields and defined through the inverse X-ray transform. We have shown that $\circ$-product is compatible with the level matching constraint and satisfies the strong constraint identically. In addition, we have shown that weakly constrained fields equipped with the $\circ$-product form a commutative ring. These features are very important for practical computations.

By associating the $\circ$-product, we have defined global $\OddZ_{\circ}$ transform corresponding to the $\OddZ$ T-duality transformation in the strongly constrained DFT. The physical degrees of freedom are given by the weakly constrained generalized metric $\cH$ and dilaton $d$. As strongly constrained DFT, we have shown that they are an $\OddZ_{\circ}$ tensor and scalar respectively. As the other weakly constrained fields, the $\cH$ and $d$ consist of strongly constrained X-ray image fields. We have introduced polarizations for the parametrization of the each strongly constrained generalized metric and dilaton.

Based on this formalism, we have constructed a gauge transformation and associated gauge invariant action for weakly constrained DFT. Using the fact that physical degrees of freedom are represented by a sum of strongly constrained generalized metrics and dilatons, we have defined the gauge transform as a sum of generalized Lie derivatives, which is the gauge transform for strongly constrained DFT, through the $\circ$-product. The gauge transformation forms a closed gauge algebra without using strong constraint. The corresponding gauge invariant action has also been represented by a sum over all possible strongly constrained DFT actions. The gauge invariance of the action is also guaranteed without strong constraint. 

In this paper, we have shown that the full relaxation of strong constraint is possible, but the relation to closed string filed theory is not clear. As we have discussed, weakly constrained DFT is defined by integrating all the massive string states, since there is no hierarchy of $\alpha'$ in weakly constrained DFT limit. In a sense, it is intriguing that there exists a well-defined effective field theory description even though the full winding modes survives. For a correct comparison to the closed string field theory, we need to consider fluctuations of closed string field including all the massive modes and integrating out all of the massive string states. However it is practically an impossible task, thus we have used the result in \cite{Hull:2009mi}, which simply ignored the string massive states in the fluctuations of string fields. However, since the naive closed string field theory result is gauge invariant without using strong constraint, it seems there is a physical implication for this theory. We have shown that the fluctuations of our action and closed string field theory are not identical, but the difference is given by the Fourier zero-modes. Also, under a certain truncation, if any, which suppresses the zero-modes only then the naive closed string field theory result reduces to the fluctuations of our construction. Therefore, our claim is that at least our construction provides an associative truncation of weakly constrained DFT or massless subsector of closed string field theory on a torus after integrating out all the string massive modes. 

%It would be interesting to investigate which limit suppress the interactions between the different null-planes. As we have discussed in \ref{}, the closed string field theory suggests that the proper binary operation between weakly constrained fields is given by projector. The projector contains ... , thus if we construct 

% From the point of view of the weakly constrained DFT with $\circ$-product, in order to reproduce the fluctuations (\ref{flucWDFT}) from closed string field theory, it seems that the $n$-string vertex should be replaced to
%
%\begin{equation}
% \lrangle{\cV_1 \cV_2 \cV_3 \cdots} \longrightarrow \lrangle{\cV_1\circ\cV_2\circ\cV_3 \cdots}\,, 
%\end{equation}
%
%where 
%\begin{equation}
%  \lrangle{\cV_1\circ\cV_2\circ\cV_3 \cdots} = \sum_{\Pi} \varphi \lrangle{\hat{\cV}_{\sPi,1}\hat{\cV}_{\sPi,2}\hat{\cV}_{\sPi,3} \cdots}
%\end{equation}
%It may define a subtheory of full closed string field theory.

%The possible advantage of the $\circ$-product of the vertex operators is that we don't need the cocycle factor, and it is always commute
%\begin{equation}
%  \cV_1 \circ \cV_2 = \cV_2 \circ \cV_1
%\end{equation}
%Because the double momenta $K_{iI}$ associated with a vertex operator $\cV_i$ are represented  on a closed null plane $\cD(X;\Pi)$
%\begin{equation}
%  K_{iI} = \ell_i \Pi^{i}{}_I \,,
%\end{equation}
%and $\Pi^i$ are null vectors, 
%\begin{equation}
%  \ell_i \Pi^{i}{}_I \ell'_j \Pi^{jI} = 0\,.
%\end{equation}
%Moreover, this feature can be extended to the three cocycle factor $\epsilon_{\alpha\beta\gamma}$.

There are a number of natural extensions of this work one could consider. We have focused only on the X-ray transform on a torus background, but these methods can be applied equally well to any homogeneous manifolds. It is not clear the meaning of the winding coordinate $\tx$ and T-duality in a homogeneous manifold. However, the inverse X-ray transform in a homogeneous manifold has already constructed with arbitrary dimensionality \cite{Helgason}, hence we may obtain a clue for these questions by studying the X-ray transform. Also, it may provide a useful framework for describing nongeometric $R$-fluxes.

The other natural generalizations of this work include weakly constrained Heterotic \cite{Hohm:2011ex}, Type II DFT \cite{TypeIIDFT1,TypeIIDFT2,TypeIIDFT3,TypeIIDFT4,TypeIIDFT5} and M-theory \cite{MDFT1,MDFT2,MDFT3,MDFT4,MDFT5,MDFT6}. Heterotic and M-theory cases are defined on extended tori $T^n$, where $n>2d$.  Since in the (inverse) X-ray transform the dimensionality of the internal torus and the closed null-plane are arbitrary, we can apply the present framework to the extended torus case. However, even if we can construct the weakly constrained M-theory, there is no known M-theory version of string field theory, thus its physical implication is not obvious. It is also interesting to consider how to include weakly constrained RR-sector in type II DFT and Yang-Mills sector in heterotic DFT.

The present formalism can be also generalized to supersymmetric cases. To this end, we need to introduce a local frame formalism and corresponding geometrical quantities such as spin-connections and curvature etc. One can easily deduce that the double-vielbein formalism in strongly constrained DFT \cite{DFTGeom1,DFTGeom2,DFTGeom3,DFTGeom4} would be easily generalized to the weakly constrained DFT case through the $\circ$-product. Also, It may be possible to generalize the fermions and supersymmetry in strongly constrained DFT straightforwardly \cite{SUSYDFT1,SUSYDFT2,SUSYDFT3}. It is interesting to investigate what kind of backgrounds which depends on both momentum and winding preserve supersymmetry by solving the Killing spinor equation. 

We have considered only torus fibre so far, but in general, we can consider the entire torus bundle. In this case, the base manifold is described by strongly constrained DFT, but the torus fibre is described in terms of weakly constrained DFT. Then the theory would take the form which is analogous to the exceptional field theory \cite{EFT1,EFT2,EFT3,EFT4}. It is also interesting to consider the equations of motion and solutions in order to study how the dynamics of winding modes on internal tori affect to the geometry of external space.

%%%%%%%%%%%%%%%%%%%%%%%%%%%%%%%%%%%%%%%%%%%%%%%%%%%%%%%
~\\~\\
\noindent{{\textbf{Acknowledgements.}}  

I would like to thank to David Berman, Chris Hull, Jeong-Hyuck Park, Felix Rudolf, Ashoke Sen and Dan Waldram for useful discussions and comments. I would also like to thank the organizers of the conference ``CERN-CKC TH Institute on Duality Symmetries in String and M-Theories'' at CERN in August 2015 where the ideas for this paper were shown.

\newpage
%%%%%%%%%%%%%%%%%%%%%%%%%%%%%%%%%%%%%%%%%%%%%%%%%%%%%%%

\end{document}